\documentclass[preprint,12pt]{elsarticle}

\usepackage{lineno}
\usepackage{latexsym}
\usepackage{amsmath,theorem}
\usepackage{amsfonts}
\usepackage{amssymb}
\usepackage{psfrag}
\usepackage{graphicx,color,subfigure}
\usepackage{soul}
\usepackage{hyperref}
\usepackage[linesnumbered,ruled,vlined]{algorithm2e}
\usepackage[normalem]{ulem}

\renewcommand{\eqref}[1]{(\ref{#1})}

\newtheorem{thm}{Theorem}

\journal{Some Journal}

\begin{document}

\begin{frontmatter}




\title{Solving an Inverse Problem for Time Series Valued Computer Simulators via Multiple Contour Estimation}


\author[1]{Pritam Ranjan}
\address[1]{Indian Institute of Management Indore, MP, India}
\author[2]{Joseph Resch}
\address[2]{University of California, Los Angeles, CA, USA}
\author[3]{Abhyuday Mandal\corref{cor1}}
\address[3]{University of Georgia, Athens, GA, USA}
\cortext[cor1]{Corresponding Author: amandal@stat.uga.edu}

\begin{abstract}
Computer simulators are often used as a substitute of complex real-life phenomena which are either expensive or infeasible to experiment with. This paper focuses on how to efficiently solve the inverse problem for an expensive to evaluate time series valued computer simulator. The research is motivated by a hydrological simulator which has to be tuned for generating realistic rainfall-runoff measurements in Athens, Georgia, USA. Assuming that the simulator returns $g(x,t)$ over $L$ time points for a given input $x$, the proposed methodology begins with a careful construction of a discretization (time-) point set (DPS) of size $k \ll L$, achieved by adopting a regression spline approximation of the target response series at $k$ optimal knots locations $\{t_1^*, t_2^*,...,t_k^*\}$. Subsequently, we solve $k$ scalar valued inverse problems for simulator $g(x,t_j^*)$ via the contour estimation method. {The proposed approach, {named MSCE,} also facilitates the uncertainty quantification of the inverse solution.} {Extensive simulation study is used to demonstrate} the performance comparison of the proposed method with the popular competitors for several test-function based computer simulators and a real-life rainfall-runoff measurement model. 
\end{abstract}


\end{frontmatter}


\section{Introduction}
\label{S:1}

Complex physical experiments are frequently expensive and impractical to perform. The growth in computing power during modern times offers an alternative to carry out such experiments via computer simulation models, such as dynamic traffic patterns of a metropolitan intersection, energy harvesting via wind farms and tidal turbines, quantification of volcanic hazards, hydrological behaviors of an ecosystem, the spread of a wildfire, weather modeling, formation of galaxies, and so on ({Kaufman et al. (2008), Bayarri et al. (2009), Mandal et al. (2009), Vernon et al. (2010), Ranjan et al. (2011),  Bingham et al. (2014), Gration and Wilkinson (2019), Kennedy et al. (2020), Krityakierne and Baowan (2020), Oberpriller et al. (2021), Lukemire et al. (2021)}). Realistic computer simulators of complex physical, engineering and sociological phenomena are often computationally expensive to run, and thus innovative design and analysis techniques have to be developed for deeper understanding of the process.

Over the last three decades, a plethora of innovative methodologies on computer experiments have been developed in the statistics and engineering literature. Some of the seminal papers focus on, the emulation of simulator response via Gaussian Process (GP) models (Sacks et al. 1989),  space-filling designs for building good surrogate models to emulate deterministic simulator outputs (Johnson, et al. 1990), sequential design approach via a merit based criterion called the expected improvement for global optimization of an expensive to evaluate simulator (Jones et al. 1998), a Bayesian approach for the emulation of simulator models in the presence of calibration parameters (Kennedy and O'Hagan 2002), treed-GP for the emulation of non-stationary simulators (Gramacy and Lee 2008), and {localized GP models (Gramacy 2016)}. For a detailed discussion on methodological development on this topic, see Santner et al. (2003), Fang et al. (2006), Rasmussen and Williams (2006) and Gramacy (2020).

In this paper we focus on solving an inverse problem for expensive to evaluate computer simulator which produces time series outputs. Let  $g(x)=\{g(x,t_j), j=1,2,...,L\}$ be the simulator output for input $x \in \chi$, a hyper-rectangle scaled to $[0,1]^d$, where $d$ is the input dimension. The inverse solution, $S_0$, with respect to a pre-specified target $g_0=\{g_0(t_j), j=1,2,...,L\}$ refers to the set of inputs $x$ that generate $g_0$, i.e.,
$$S_0 = \{x \in \chi: g(x, t_j) = g_0(t_j), j=1,2,...,L\}.$$ 
The application that motivated this study comes from a hydrological simulation model which predicts the rate of rainfall-runoff and sediment yield for a windrow composting pad (Duncan et al., 2013). Here, the objective is to find the inputs of the hydrological model that generates outputs as close as possible to the real data collected from a watershed from the Bioconversion center at the University of Georgia, Athens, USA.

Inverse problem for expensive to evaluate \emph{scalar valued} simulators has been extensively investigated in the past few years (e.g., Oakley (2004), Ranjan et al. (2008), Bichon et al. (2008), Picheny et al. (2010), Bect et al. (2012), Roy and Notz (2014), Jala et al. (2016), Azzimonti et al. (2021), Cao et al. (2021), Cole et al. (2021)). A closely related research topic is referred to as the estimation of calibration parameters, where the computer simulator takes two types of inputs, controllable design variables and fixed but unknown calibration parameters. Kennedy and O'Hagan (2002) proposed a Bayesian framework that accounts for the two types of inputs, and models a potential systematic discrepancy between the observed field data and the simulator response. This model received significant attention in both computer experiments and engineering literature, for instance, Tuo and Wu (2015), Pratola et al. (2013), Brown and Hund (2018), Perdikaris and Karniadakis (2016), Perrin (2020).

For simulators with \emph{{time series}} response and only controllable inputs, the inverse problem literature include Vernon et al. (2010), Ranjan et al. (2016), Zhang et al. (2019), Bhattacharjee et al. (2019), and Toscano-Palmerin and Frazier (2022). We focus on this setting. Vernon et al. (2010) developed an innovative history matching (HM) algorithm for solving the inverse problem of a galaxy formation model called GALFORM. This is multi-stage sampling strategy, which intelligently eliminates the implausible points from the input space and returns a set of plausible candidates.  The main idea was to first select a handful of time-points from the target response series (referred to as the discretization point set (DPS) = $\{t_1^*, t_2^*,...,t_k^*\}$, where $1 \le t_1^*<\cdots < t_k^* \le L$), and then optimize a joint discrepancy criterion (called the implausibility function) between the target and predicted response from the GP surrogates of $g(x,t_j^*)$ $-$ the scalar-projection of the process at DPS locations $t_j^*$, for $j=1,...,k$. It turns out that the HM algorithm requires too many simulator runs, and for expensive to evaluate computer simulators, this approach would be un-affordable. Recently, Bhattacharjee et al. (2019) proposed a small modification in the sampling strategy of the HM algorithm which reduced the required simulator runs without compromising the accuracy of the inverse solution. 

In a simplified approach to such an inverse problem, Ranjan et al. (2016)  introduced a new pseudo scalar-valued simulator $w(x)=\|g(x)-g_0\|$ and then find the minimizer of $w(x)$ using a global optimization method $-$ build GP surrogate for $w(x)$ coupled with sequential design techniques via the expected improvement (EI) criterion developed by Jones et al. (1998). In the same spirit, Zhang et al. (2019) minimized $w(x) = \|g(x)-g_0\|$, however, instead of fitting a scalar-valued GP surrogate of $w(x)$, the authors used a singular value decomposition (SVD)-based GP surrogate (Higdon et al., 2008) for $g(x)$ and developed a saddlepoint approximation of the EI expression of Jones et al. (1998), referred to as the saEI approach.

{In this paper, we proposed a new MSCE method for solving an inverse problem for time-series valued computer simulators.The proposed approach has two key components. Inspired by the HM algorithm, we first discretize the target response series at DPS. However, Vernon et al. (2010) and Bhattacharjee et al. (2019) used an ad-hoc method (or a subjective expert opinion) for choosing the DPS. We suggest a more formal approach by fitting a regression spline to the target series $g_0$ and then identify the desired DPS as the optimal knot locations. We investigated both the sequential search and the simultaneous search methods for finding optimal knots. Then, we solve $k$ scalar-valued inverse problems, i.e., estimate
$$ S_j = \{x\in \chi : g(x,t_j^*) = g_0(t_j^*) \}, \, j=1,2,...,k.$$
Finding $S_j$ is essentially a contour estimation problem, as in Ranjan et al., (2008). As per our knowledge the existing literature on inverse problems for time series valued simulators (e.g., Vernon et al. (2010), Ranjan et al. (2016) and Zhang et al. (2019)) use the global minimization criterion by Jones et al. (1998). In this paper, we propose using the contour estimation EI criterion for iteratively solving these $k$ scalar-valued inverse problems.} At the end, the inverse solution of the underlying dynamic simulator is obtained by taking the intersection of all scalar-valued inverse solutions, {which is further used to quantify the uncertainty associated with the estimated inverse solution.} Theoretical result which establishes the estimation of the desired inverse solution is also presented. {Extensive simulation studies have been used to demonstrate that the proposed approach is more accurate and reliable than its competitors. The results are compared with those of modified HM algorithm (Bhattacharjee et al. 2019), scalarization method (Ranjan et al. 2016), and saEI method (Zhang et al. 2019).}

The remaining sections are outlined as follows. Section 2 reviews the concepts integral to the proposed method and the competing approaches, i.e., the scalarization method by Ranjan et al. (2016), the HM approach proposed by Vernon el al. (2010) with modification in Bhattacharjee et al. (2019), {and the saEI method by Zhang et al. (2019)}. Section 3 provides the elements of the proposed multiple scalar-valued contour estimation (MSCE) method, {uncertainty quantification of the inverse solution}, and thorough implementation details of the steps. In Section 4, we {present simulation studies to} establish the superiority of the proposed method via three test functions based simulator. Section~5 discusses the real-life motivating hydrological-simulation application. We provide some concluding remarks in Section 6.

\section{Review of Existing Methodology}
\label{S:2}

In this section, the existing methods that set precedence for this paper are presented. We briefly review GP-based models used as surrogates of the simulator outputs and the EI criterion for choosing the follow-up trials in the sequential design framework. Subsequently, the scalarization method by Ranjan et al. (2016), HM algorithm of Vernon et al. (2010) and Bhattacharjee et al. (2019), and saEI method by Zhang et al. (2019) are also reviewed.

\subsection{Gaussian Process-based Models}
The evaluation of a computer simulator for complex phenomena can often be computationally expensive, and hence the emulation via a statistical surrogate becomes much more practical. Sacks et al. (1989) presents a  GP model as a useful surrogate of deterministic simulator output. For a set of input-output combinations, a stationary GP model, also called as the ordinary Kriging, assumes:
\begin{eqnarray*}
	y(x_i) = \mu + Z(x_i), \hspace{0.5in} i = 1, \dots, n,
\end{eqnarray*}
where $\mu$ is the mean and $Z(x_i)$ is a GP with $E(Z(x_i)) = 0$ and a covariance structure of $Cov(Z(x_i), Z(x_j)) = \sigma^2R(\theta; x_i, x_j)$. There are several popular choices of $R(\cdot,\cdot)$, e.g., Gaussian correlation, power-exponential family, and Mat\'ern correlation. The power-exponential correlation structure will have the $(i,j)^{\mbox{th}}$ term $R_{ij}(\theta)$ as:
\begin{eqnarray}\label{eq:1}
R(Z(x_i), Z(x_j)) = \prod_{k = 1}^{d} \exp\Bigg\{-\theta_k \mid x_{ik} - x_{jk} \mid ^{p_k} \Bigg\} \hspace{0.2in} \mbox{for all }  i,j,
\end{eqnarray}
where $0 < p_k \leq 2$ are smoothness parameters and $\theta_k$ measure the correlation strength. In this paper, we assume power exponential correlation with $p_k = 1.95$ (for numerical stability and smoothness). The best linear unbiased predictor for the response at any unsampled point $x^*$ is given by:
\begin{eqnarray}\label{eq:yhat}
\hat{y}(x^*) = \hat{\mu} + r(x^*)^T R_n^{-1}(y-1_n \hat{\mu}),
\end{eqnarray}
where $r(x^*) = \Big[ \mbox{corr} (z(x^*), z(x_1)), \dots, \mbox{corr} (z(x^*), z(x_n))\Big]^{\tt T}$, $R_n$ is the $n\times n$ correlation matrix with elements $R_{ij}$ (as in Equation (\ref{eq:1})), and the prediction uncertainty is quantified by
\begin{eqnarray}\label{eq:2}
s^2(x^*) = \hat{\sigma}^2 \bigg(1-r(x^*)^T R_n^{-1}r(x^*) \bigg).
\end{eqnarray}

The flexibility of the correlation structure, and the closed form expressions for mean prediction and associated uncertainty makes the GP model a popular surrogate for complex computer model outputs. Throughout this paper, the \textit{R} package {\tt GPfit} (MacDonald et al., 2015) has been used to fit the basic scalar-valued GP models. 

{Fitting a GP model requires numerous inverse calculations of size $n\times n$ each, which becomes computationally daunting as $n$ increases and particularly for simulation studies when the entire exercise has to be repeated thousands of times. Gramacy (2016) developed an \textit{R} package called {\tt laGP} -- a local approximate GP (laGP) model for large data sets. The main idea is to fit local GP model for prediction at any given point in the input space. The process of finding the local set of size $m (\ll n)$ starts with a $k$-nearest neighbor set around the point of interest, and then selecting the remaining $m-k$ points guided by a model-based criterion.  Finally, the prediction at the point of interest is obtained using the GP model built on this local neighborhood set of size $m$.  See  Gramacy and Apley (2015) for methodological details. In this paper, if $n>50$, {\tt laGP} package has been used for all GP model fittings within the simulation exercises. For $n\le 50$, we used the {\tt GPfit} package.}

\subsection{Sequential Design}
It has been established on many occasions that sequential designs outperform its competitors for finding a pre-specified feature of interest, e.g., the global minimum or the inverse solution, for a computationally intensive deterministic scalar-valued computer simulator (Jones et al. (1998), Santner et al. (2003), Forrester et al. (2007),  Ranjan et al. (2008), Picheny et al. (2010),  Zhang et al. (2019), and Gramacy (2020)). The basic framework consists of three key steps, finding a good initial design, fitting the statistical surrogate and choosing the follow-up trials by optimizing a merit based criterion (EI is the most popular one).

In computer experiments, the popular choice for an initial design includes a space-filling design such as a maximin Latin hypercube design (LHD) ({Morris and Mitchell (1995), Wang et al. (2021)}), a maximum projection LHD (Joseph et al., 2015), uniform design, and orthogonal array based LHD ({Wang et al., 2020}). Once an initial design is chosen, the responses are generated by evaluating the simulator at each input. A surrogate model is then fitted to the training data $(x_i, y_i), i=1,2,...,n$. We use the GP / laGP model (detailed in Section~2.1) for this purpose. After which, a sequential design criterion such as EI is evaluated  over the entire input space to find the input $x_{new}$ $-$ the maximizer of EI (see Jones et al. (1998) and Bingham et al. (2014) for details).  The $x_{new}$ and corresponding true simulator response are augmented to the training set (i.e., $n=n+1$). The surrogate (GP model) is refitted to this augmented training set. The iterative process of optimizing EI to choose $x_{new}$ and refitting the surrogate to the augmented data, is repeated until the total budget of $N$ points is exhausted. The feature of interest (e.g., the global optimum or the inverse solution)  would be extracted from the final surrogate fit.

\subsection{Inverse Problem via Scalarization}

Ranjan et al. (2016) assumed $w(x)=\|g(x)-g_0\|$ to be the output of a new scalarized simulator which is expensive to evaluate, and thus the popular sequential approach by Jones et al. (1998) was applied to find the global minimum. That is, a GP model (Section~2.1) is used to emulate the scalar-valued response $w(x)$, and the EI criterion by Jones et al. (1998) dictates how to choose the follow-up points. 
%
%
Note that in Section~2.1, $y(x)$ denotes a scalar simulator response, whereas in this section, we denote $w(x)$ as the scalar response. The EI criterion, as per the Gaussian predictive distribution with mean and variance given by (\ref{eq:yhat}) and (\ref{eq:2}),  has a closed form expression 
$$ E[I(x)] = (w_{min}-\hat{w}(x)) \Phi\left(\frac{w_{min}-\hat{w}(x)}{s(x)}\right) + s(x)\phi\left(\frac{w_{min}-\hat{w}(x)}{s(x)}\right),$$
where $\phi(\cdot)$ and $\Phi(\cdot)$ are the normal probability density function (pdf) and cumulative distribution function (cdf) respectively. 

The EI based approach has gained immense popularity because it facilitates a balance between the exploration and exploitation, which further implies that the entire input space is explored thoroughly and hence eventually all global minima would be found. That is, if there are more than one solution of the inverse problem, then the EI-based approach would be able to detect them. Finally, the inverse solution is obtained by minimizing the responses over the training data or the predicted response over a dense set via the final fitted surrogate.

\subsection{EI Criterion for Contour Estimation}

For a scalar valued deterministic computer simulator,  Ranjan et al. (2008) developed an EI criterion for estimating the inputs that lead to a pre-specified target response $y(x)=a$. The proposed improvement function is given by 
\begin{equation*}\label{Eq.EI_contour1}
I(x^{*}) = \epsilon^{2}(x^{*}) - \mbox{min}\Big[ \{ y(x^{*}) - a\}^{2}, \epsilon^{2}(x^{*}) \Big].
\end{equation*}
where $\epsilon(x^{*}) = \alpha s(x^{*})$ for a positive constant $\alpha$ (e.g., $\alpha = 0.67$ corresponds to 50\% confidence, and $\alpha=1.96$ represents 95\% level of confidence, under normality), $s(x^*)$ is defined in (\ref{eq:2}), and $a$ is the pre-specified target response.  Hence, the EI value (which is simply the expected value of the improvement function under the predictive distribution $y(x^*) \sim N(\hat{y}(x^*), s^2(x^*))$) is:
\begin{eqnarray}\label{Eq.EI_contour2}
E[I(x^{*})] =& \Big[\epsilon^{2}(x^{*}) - \{ \hat{y}(x^{*}) - a \}^{2} \Big] \Big\{\Phi(u_{2}) - \Phi(u_{1})\Big\} \nonumber \\
&+ s^{2}(x^{*})\Big[ \{u_{2} \phi(u_{2}) - u_{1} \phi(u_{1})\} - \{\Phi(u_{2}) - \Phi(u_{1})\}\Big] \nonumber \\
&+ 2\Big\{ \hat{y}(x^{*}) - a \Big\}s(x^{*}) \Big\{\phi(u_{2}) - \phi(u_{1})\Big\},
\end{eqnarray}
where $u_{1} = [a - \hat{y}(x^{*}) - \epsilon(x^{*})] / s(x^{*})$, and $u_{2} = [a - \hat{y}(x^{*}) + \epsilon(x^{*})] / s(x^{*})$.   

Similar to the EI in Jones et al. (1998), this EI criterion also facilitates the balance between local and global search. {In other words, all pieces of the contours are expected to be detected eventually.}

\subsection{History Matching for the Inverse Problem} \label{hm_via_sd}

HM approach was developed by Vernon et al. (2010), and was subsequently modified by Bhattacharjee et al. (2019) to solve the inverse problem for a {time series} valued computer simulator.  The HM approach starts by selecting a handful of time-points \{$t_1^*, t_2^*, \dots, t_{k}^*$\}, which are referred to as a DPS {and has size $k$, which is significantly smaller than $L$, the total length of the response series.}  The said approach uses the simulator outputs at only the DPS time-points and approximates the desired inverse solution by eliminating the set of implausible points from the input space via an innovative criterion called the implausibility function.

The HM algorithm is implemented via a multi-stage sampling technique. First a large space-filling initial design $\{x_1, x_2, ..., x_n \}$ is used to evaluate the {time series} valued simulator, and extract the scalar projections of the input-output training set at the DPS locations. {Subsequently, the algorithm iterates between the following four steps:
	
	\begin{enumerate}
		\item For $j=1,2,...,k$, fit $k$ scalar-valued GP surrogates to $\{(x_i, g(x_i, t_j^*)), i=1,2,...,n\}$, where $n$ is the size of the training set.
		\item Evaluate a criterion called the implausibility function over a large test set. For each $j = 1, \dots, k$, the implausibility criterion is defined as
		\begin{equation*}
		IM_{j}(x) = \frac{\mid\hat{g}(x, t_j^*) - g_0(t_j^*)\mid}{s_{t_j^*}(x)},
		\end{equation*}
		where $\hat{g}(x, t_j^*)$ is the predicted response derived from the GP surrogate corresponding to the simulator response at time point $t_j^*$ and $s_{t_j^*}(x)$ is the associated uncertainty. From the test set, points are deemed implausible if $IM_{max}(x) > c$, where $c$ is the pre-determined cutoff chosen in an ad-hoc manner and 
		$$	IM_{max}(x) = \mbox{max}\{IM_{1}(x), IM_{2}(x), \dots, IM_{k}(x)\}.$$ 
		Points in the complement set are said to be plausible.
		\item Select the plausible design points, augment it to the training set, and go to Step 1.
		\item At the end of the procedure, the approximate inverse solution is extracted from the training set or  from the predicted response over a dense set via the final surrogate.  
	\end{enumerate}	
}

Bhattacharjee et al. (2019) recommended a modification in the HM algorithm and used a small initial design as per the popular $n_0=10\cdot d$ rule-of-thumb  {(Loeppky et al., 2009, Harari et al., 2018)} as compared to a large initial design. This helped in achieving the  desired accuracy of the inverse solution with significantly fewer runs. { However, the size of the training set in  Bhattacharjee et al. (2019) can still become very large very fast because the algorithms recommends choosing all plausible points in Step~3. For instance, their (Matlab-simulink) hydrological model example required 461 simulator runs for estimating the inverse solution. In this paper, we implement a sub-sampling strategy via clustering and then select only the cluster centers instead of all plausible points. This will ensure that the input space is explored thoroughly with much fewer training points. We follow this two-fold modified HM algorithm for all simulations.}

\subsection{Saddlepoint Approximation-based EI} \label{saEI_section}

{Zhang et al. (2019) used SVD-based GP model originally developed by Higdon et al. (2008) for fitting a surrogate to the {time series} output $g(x)$ of a computer simulator. Although slightly more complicated, but here also, the predicted mean response and the associated uncertainty (i.e., mean square error) have closed form expressions. Subsequently, the authors applied the EI criterion in Jones et al. (1998) to $w(x)=\|g(x)-g_0\|$, i.e., 
$$ E[I(x)] = E[(w_{min}-w(x))_+|Data], $$
however, the expectation had to be computed with respect to the SVD-GP - the surrogate model for $g(x)$. Here the authors proposed a saddlepoint approximation for computing $E[I(x)]$. They also developed an \textit{R} package called {\tt DynamicGP} which implements this methodology. The usage of the most important function called \texttt{saEI} is shown as follows:}
\begin{verbatim}
       saEIout = saEI(xi,yi,yobs,nadd,candei,candest,func,...,
                   nthread=4,clutype="PSOCK") 
\end{verbatim}
{where \texttt{xi} and \texttt{yi} denote the initial training data, \texttt{yobs} is the target response, \texttt{nadd} is the number of follow-up points to be added, \texttt{candei, candest} are the test sets for optimizing the saEI criterion and extracting the inverse solution respectively. Since the SVD-GP model fitting and saddlepoint approximation calculations are computationally intensive, parallel computing environment can also be used via specifying the number of threads (\texttt{nthread}) and cluster type (\texttt{clutype}).
}

\section{Proposed Methodology}

Most of the existing methodologies to solve the inverse problem for simulator with {time series} response use the global minimization criterion by Jones et al. (1998). We propose a methodology that is based on the usage of scalar-valued contour estimation criterion by Ranjan et al. (2008) for the inverse problem under a limited budget constraint. 

Similar to the HM algorithm, we discretize the simulator response at a DPS of size $k (\ll L)$ that aims to capture the important features of the target response series.  However, instead of choosing the DPS via a subjective judgement, we propose using a systematic construction approach via regression spline approximation of the target series $g_0$. Subsequently, we propose to iteratively solve the $k$ scalar-valued inverse problems using the efficient contour estimation method (outlined in Sections~2.2 and 2.4). Finally, the desired inverse solution is obtained by taking the intersection of these $k$ sets of inverse solutions. There are several parts of the proposed methodology that requires detailed discussion.

\textbf{Construction of DPS:}  Fit a (regression) cubic spline function to the target response series $g_0$ and then use the set of knot locations as the DPS. However, finding optimal set of knots is a classical yet challenging problem.  

Two obvious approaches to address this issue are ``simultaneous search" and ``sequential search".  The ``simultaneous search" finds the best $k$-knot combination by simultaneously searching the $k$-dimensional time-point grid with  ${L \choose k}$ options and optimize a goodness of fit criterion like mean square error (MSE) for the fitted spline approximation. Subsequently,  the optimal value of $k$, and the corresponding set of knots, can be obtained using elbow method, where the MSE is plotted  against the number of knots and the objective is to identify the elbow of the plot.  

{On the other hand, the alternative} ``sequential search" follows a greedy approach for constructing the DPS. The idea is similar to the construction of a regression tree, where the split-points are essentially the knot locations. That is, we start with no knots, and find the best location for the first knot by minimizing the overall MSE as per the spline regression fit. The optimal location for the second knot is found by fixing the first knot location. Continuing further in this manner, the search for optimal location for the $j$-th knot assumes that the optimal location of the previous ($j-1$) knots are known. Finally, the optimal number of knots are found using the elbow method. For implementation, the R package {\tt splines} is called upon for this purpose while the command {\tt bs()} is used for finding B-spline basis functions in the linear model environment.

We quickly illustrate the sequential search scheme by applying it to a test function. Suppose the simulator outputs are generated via Easom function (Michalewicz, 1996),
\begin{equation*}
g(x,t_j) = \cos(x_1)\cos(x_2) \exp\big\{-(x_1-\pi t_j)^2-(x_2-\pi)^2)\big\},
\end{equation*}
where $t_j$ are $L$ equidistant time points scaled in $[0, 1]$ for $j = 1, \dots, L = 200$, and the input space is scaled to $(x_1, x_2) \in [0, 1]^2$. We select the target response $g_0$ corresponding to the input set $x_0 = (0.8, 0.2)$. Pretending that $x_0$ is unknown, the objective of the inverse problem would be to find $x=(x_1,x_2)$ such that $g(x,t_j) \approx g_0(t_j)$ for all $j=1,\ldots,200$.

The first element of the DPS is obtained by minimizing the MSE of the cubic spline {fitted to} the target response over each of the possible 200 time points as the sole knot. We found the optimal first knot at time point $t_1^*=145$. Keeping the knot at time point $t_1^*=145$ fixed, we repeated the process and tried the remaining 199 options, and found the second optimal knot at time point $t_2^*=37$. The process continued, and the locations of ten optimal knot are $\{145, 37, 132, 47, 120, 55, 113, 63, 104, 174\}$ (see Figure~1). 

\begin{figure}[h!]\centering
	\includegraphics[scale=0.6]{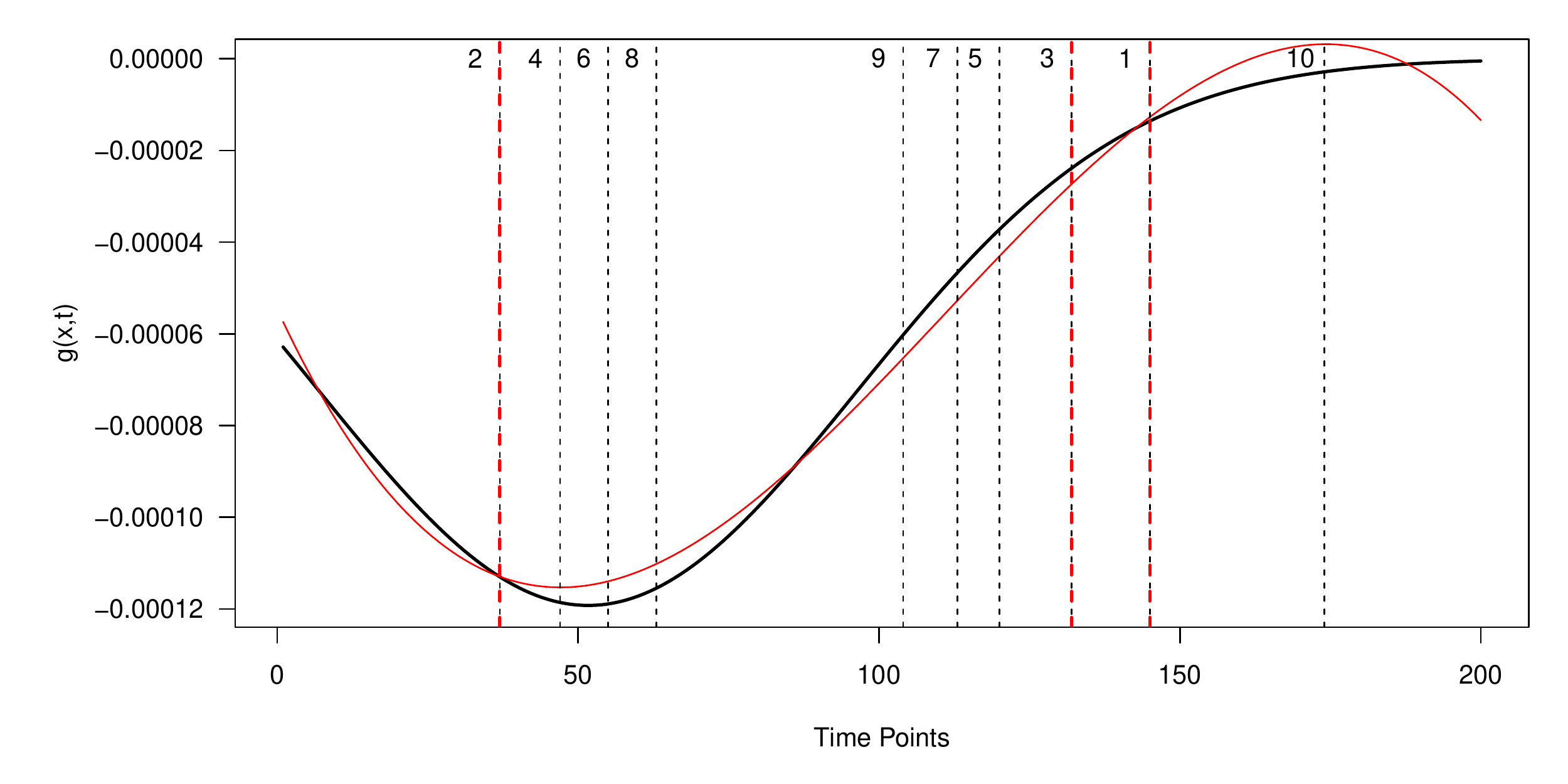} 
	\caption{Easom function: Black solid curve shows $g(x_0)$. The vertical dashed lines depict the ordered positioning of optimal knots for fitting cubic splines to the {time series}. Red curve shows the reconstructed response using DPS as knots.}
	\label{Fig: easom-knots}
\end{figure}

In Figure~1, we have illustrated the sequential selection of 10 knots, however, in reality, the required number of knots may be different.  The elbow plot method investigates the relationship between MSE and the number of knots, and finds the elbow of the plot, i.e., the second derivative reaches a positive value. This would allow for a good fit while maintaining the efficiency of the knots used. Figure~2 shows the corresponding ``MSE vs. the number of knots function" plot for the Easom function. In this case, the elbow cutoff is $3$. That is, the recommended discretization-point-set (DPS) for this time series response would be $\{145, 37, 132\}$.

\begin{figure}[h!]\centering
	\includegraphics[scale=0.7]{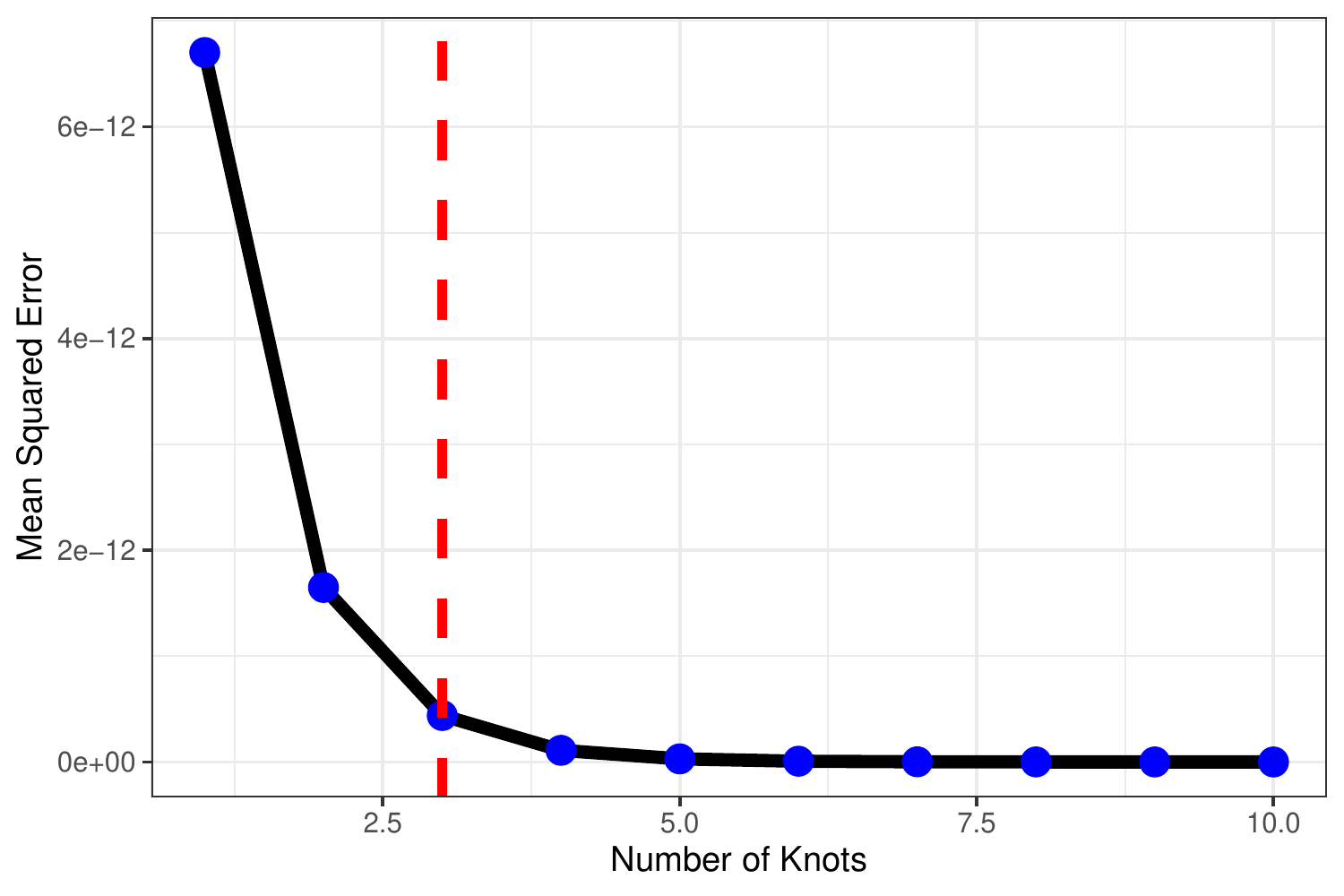} 
	\caption{Easom function: ``Mean squared error versus the number of knots" for 10 knots for spline regression added sequentially one at-a-time. }
\end{figure}

\emph{Remark 1: Computational Cost:} Although more accurate than its competitors, the simultaneous search is computationally too expensive {(dimension of the search space, ${L \choose k}$, grows exponentially for large $k$)}. As a result, it may be preferred to settle with a slightly sub-optimal (but computationally tractable) set of knots, perhaps via minimizing the goodness of fit criterion (e.g., MSE) over a randomly chosen large subset of the $L^k$ grid. Alternatively, one can use the sequential search method discussed above. For all inverse problem estimation examples considered in this paper, we have used the sequential search method for constructing DPS. Of course, in some cases, such a sub-optimal method may require a few more discretization points in the DPS to reach the desired accuracy level as compared to the ``simultaneous search" method. {In the Appendix we presented} a more detailed comparison of the computational costs.

{For a quick reference, we compare the accuracy of the two search methods for Easom test function (see Figure~\ref{Fig: easom-knots}), where the target series is generated using  $x_0 = (0.8, 0.2)$  with additional Gaussian noise. We fitted cubic-spline regression model to the target series with $j$ knots identified using the two search methods. For finding optimal DPS of size $j$ under the simultaneous search method,  we followed a computationally cheaper approximation and randomly selected $200\cdot j$ candidate points instead of fitting ${200 \choose j}$ MLR models. Figure~\ref{Fig: hs-dps-construction-cost} compares the log(MSE) of the fitted models.
	\begin{figure}[h!]\centering
		\includegraphics[scale=0.55]{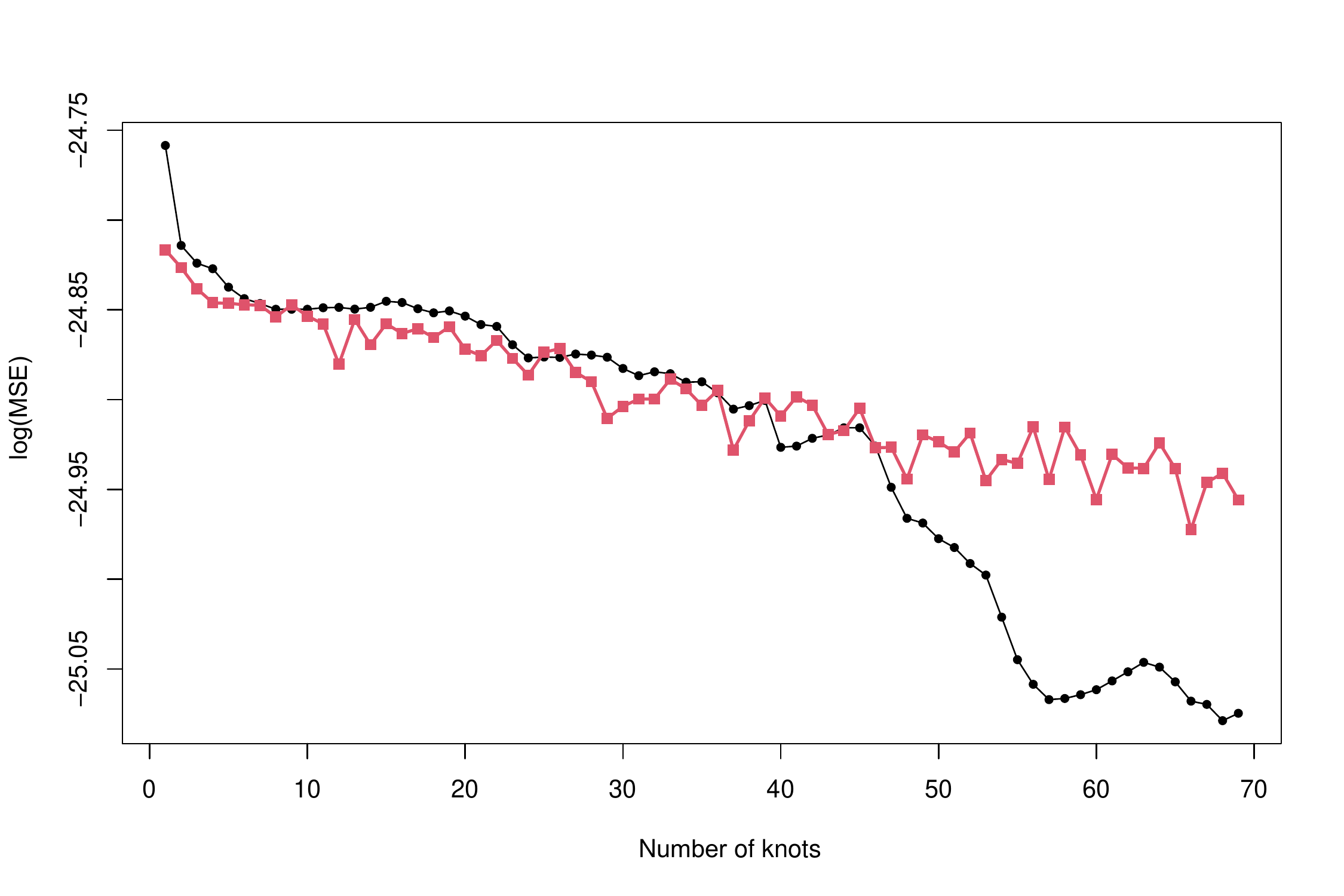}
		\caption{Easom Function: log(MSE) comparison of splines fitted to the target response, with optimal knots found using two methods: simultaneous search (red squares) and sequential search (black dots).} 
		\label{Fig: hs-dps-construction-cost}
	\end{figure}
	
It is clear from Figure~\ref{Fig: hs-dps-construction-cost} that the simultaneous search method (although computationally more expensive) provides slightly more accurate set of knots (i.e., DPS) for the initial values of $j$, but, eventually the sequential search scheme exhibits its superior performance. See Appendix for more comprehensive computational cost comparison.}

\textbf{Multiple Scalar-valued Contour Estimation (MSCE):} After finding a reasonable DPS, we sequentially estimate $k$ scalar-valued inverse solutions $S_j = \{x \in [0,1]^d: g(x,t_j^*)=g_0(t_j^*)\}$ for $j=1,2,...,k$. Suppose our total simulator run budget is $N$, then, the process starts by  choosing an initial design of size $n_0 (<N)$ from the input space $[0,1]^d$, for which we use a maximum projection Latin hypercube design (Joseph et al., 2015). The remainder of the budget $(N-n_0)$ is equally distributed in to $k$ parts for estimating $S_j, j = 1,2,...,k$. That is, the first inverse problem would estimate $S_1$ using $n_0$-point initial design and $(N-n_0)/k$ follow-up trials chosen one at-a-time by maximizing the EI criterion  (\ref{Eq.EI_contour2}) and updating the GP surrogate iteratively. The augmented data of size $n_0 + (N-n_0)/k$ are now treated as the initial training set for the second scalar-valued inverse problem. Thus, one would estimate $S_j$ using the initial training set of size $n_0 + (j-1)(N-n_0)/k$, obtained after solving the previous ($j-1$) scalar-valued inverse problems, and $(N-n_0)/k$ follow-up trials via EI optimization.

\begin{figure}[h!]\centering
	\includegraphics[scale=0.5]{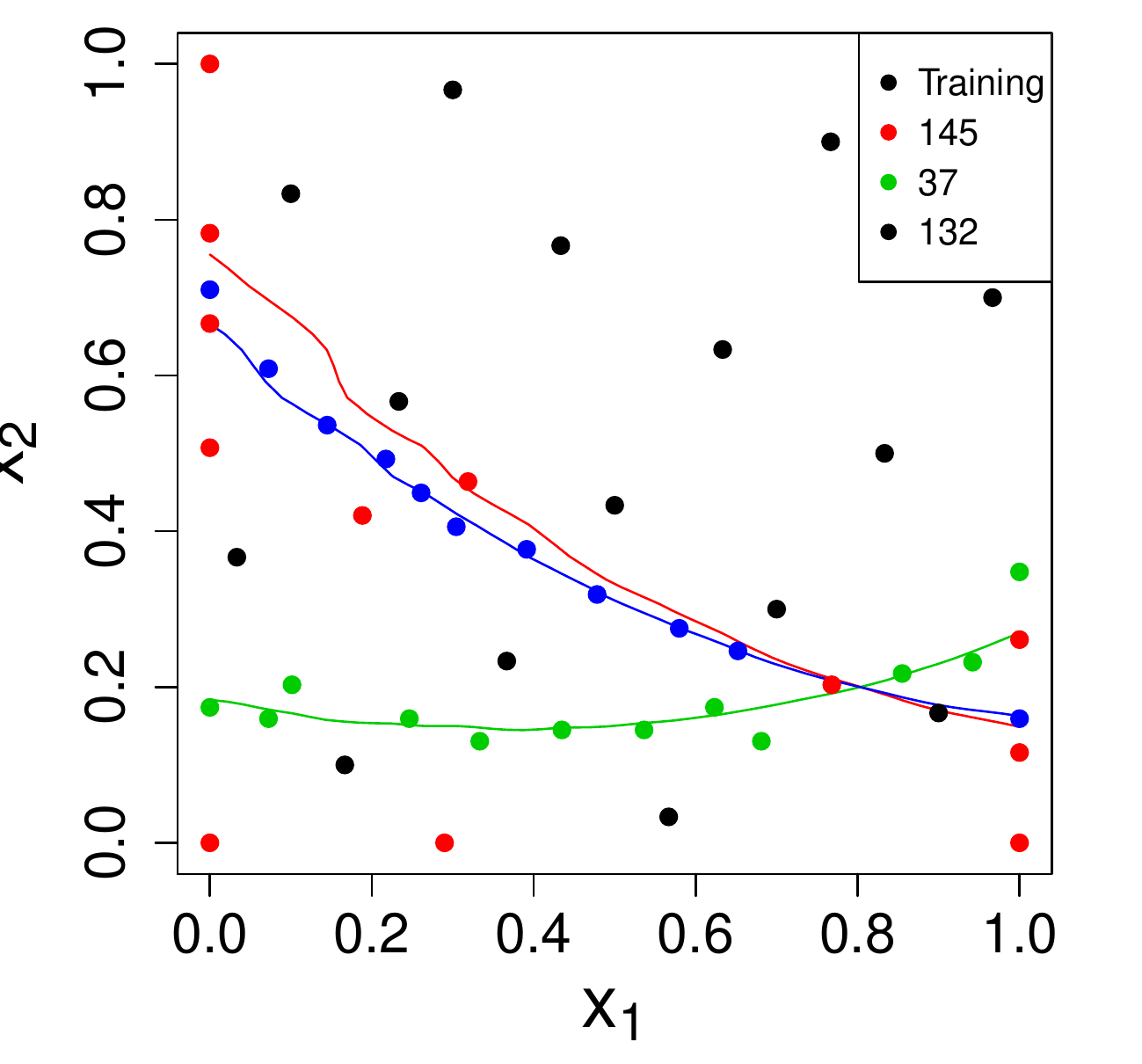}  
	\includegraphics[scale=0.6]{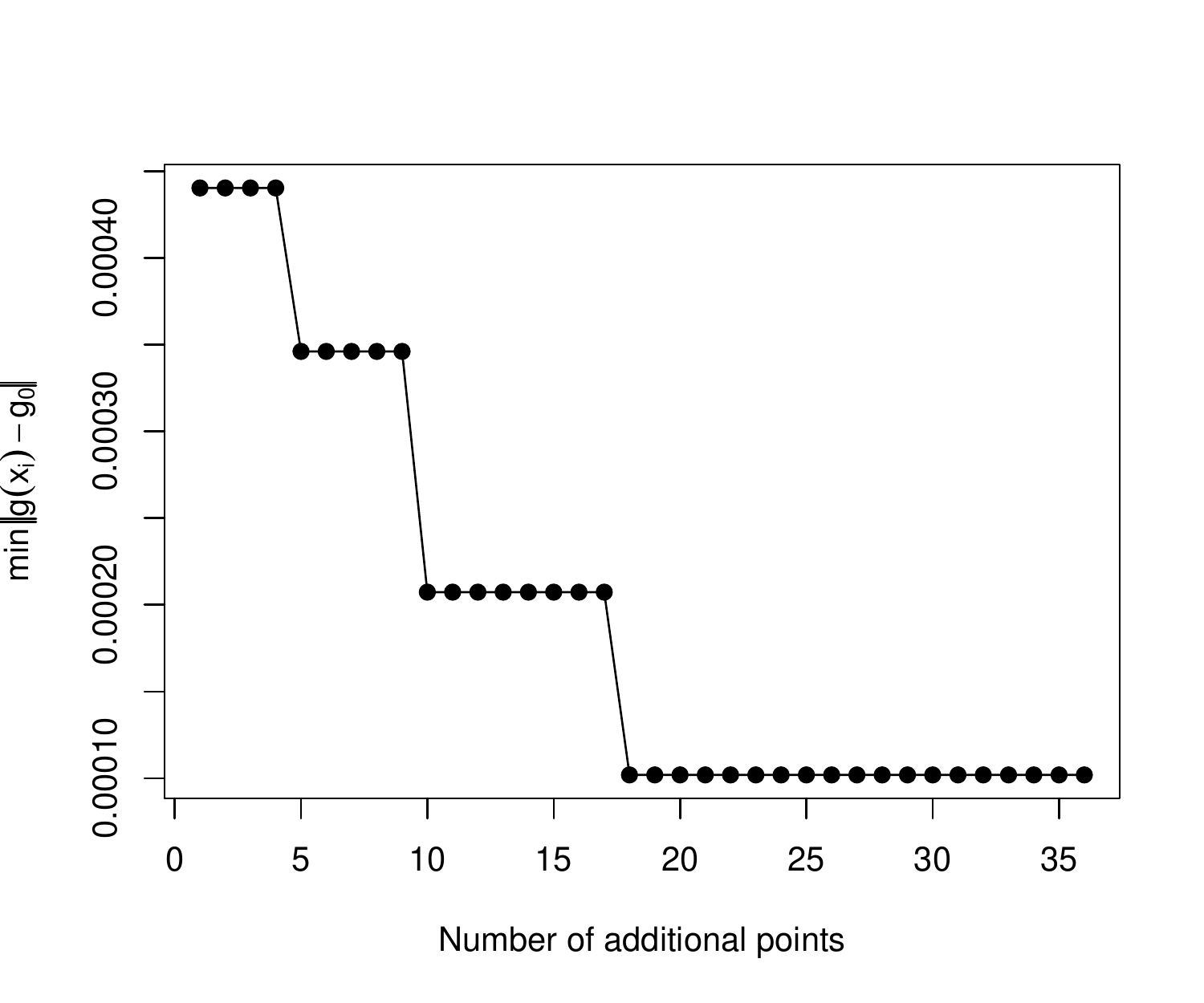}  
	\caption{Easom function: Training data is depicted by dots and the estimated contours are shown by solid curves. The black dots correspond to the initial design, whereas red, green, and blue dots represent the follow-up points obtained via EI optimization for the three scalar-valued contour estimation at $DPS =(145,37,132)$, respectively.}
	\label{fig:easom-seq-3-DPS-points}
\end{figure}

For the Easom function, since the DPS is of size three, we need to solve three scalar-valued inverse problems. We set a total training size budget of $N=50$ points and initial design of size $n_0 = 15$. The budget of follow-up points, $N - n_0 = 35$, is divided approximately evenly for the three inverse problems (i.e., $35=12+12+11$). When computing the EI criterion, we set $\alpha = 0.67$ which corresponds to $50\%$ confidence interval under normality. Furthermore, since the input space is only two-dimensional unit square, we use $5000$-point random Latin hypercube designs for maximizing the EI criteria for sequentially adding follow-up trials. The left panel of Figure~\ref{fig:easom-seq-3-DPS-points} shows the three estimated contours along with selected follow-up points corresponding to $t_1^*=145$ (in red), $t_2^*=37$ (in green) and $t_3^* = 132$ (in blue). {The right panel depicts the convergence over iterations as the follow-up points are added to the training data. The progress is measured by the minimum value of $\|g(x_i)-g_0\|$ over the training data.}

From Figure~\ref{fig:easom-seq-3-DPS-points}, it is clear that for the first contour estimation, more follow-up points focus on global exploration for better overall understanding of the process as compared to the local search for accuracy enhancement of the contour estimate.  For the second and third contour estimations the follow-up points tend to focus more and more on the local search. { The second panel of Figure~\ref{fig:easom-seq-3-DPS-points} shows that a good approximation of the inverse solution was obtained after a few additional points were added for the second contour estimation problem. }

\begin{figure}[h!]\centering
	\subfigure[DPS-37-145-132]{\includegraphics[scale=0.35]{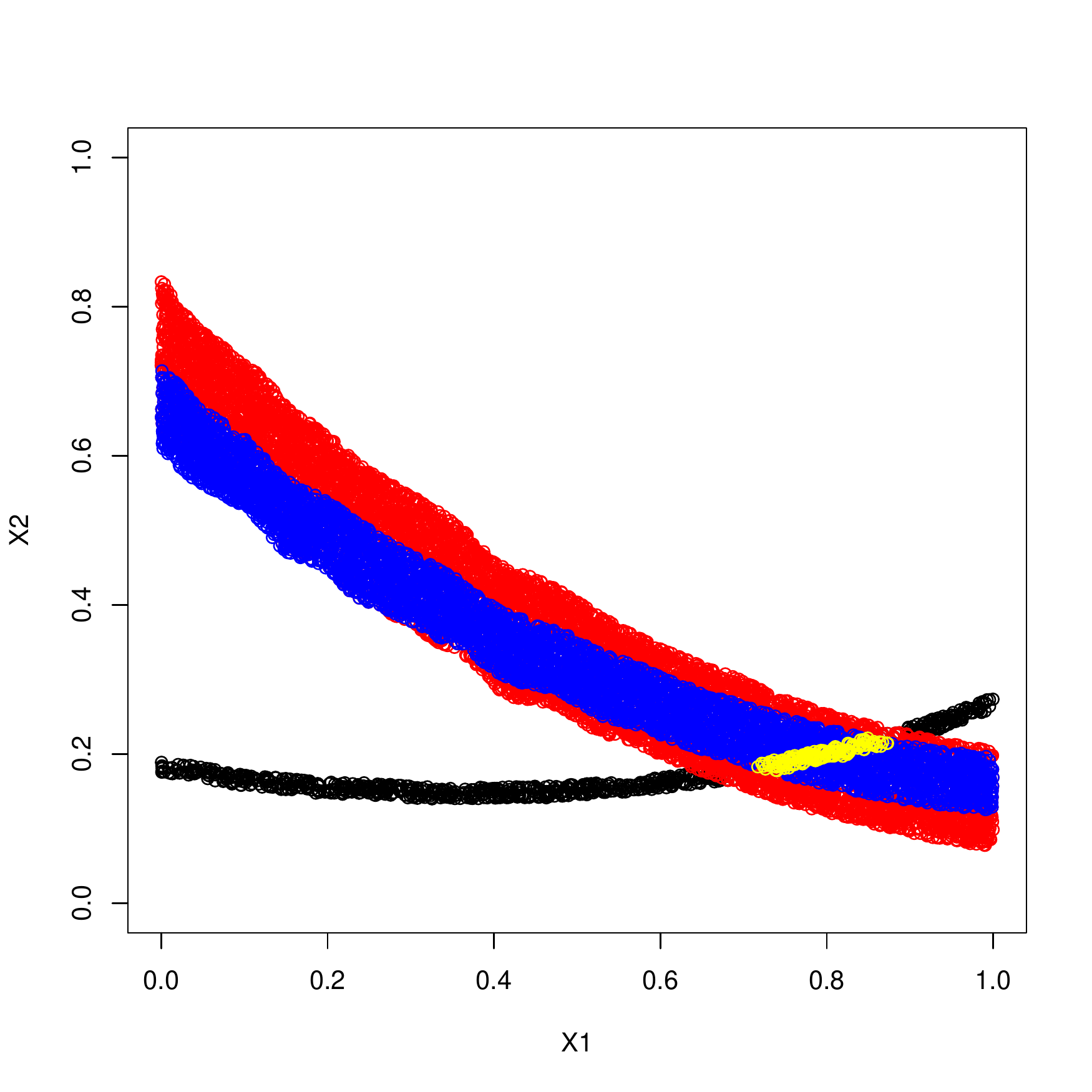}}
	\subfigure[DPS-132-145-37]{\includegraphics[scale=0.35]{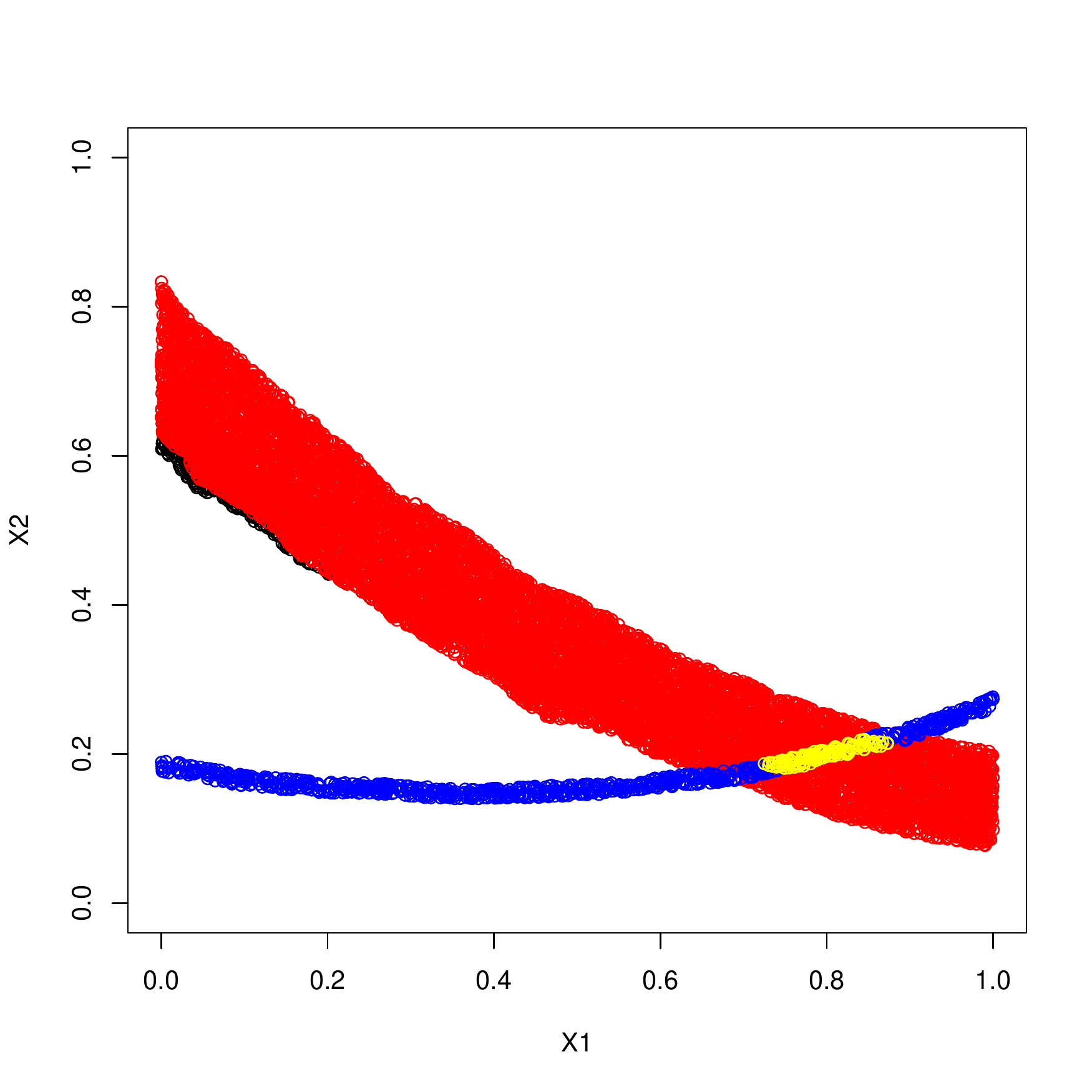}}
	\subfigure[DPS-145-132-37]{\includegraphics[scale=0.35]{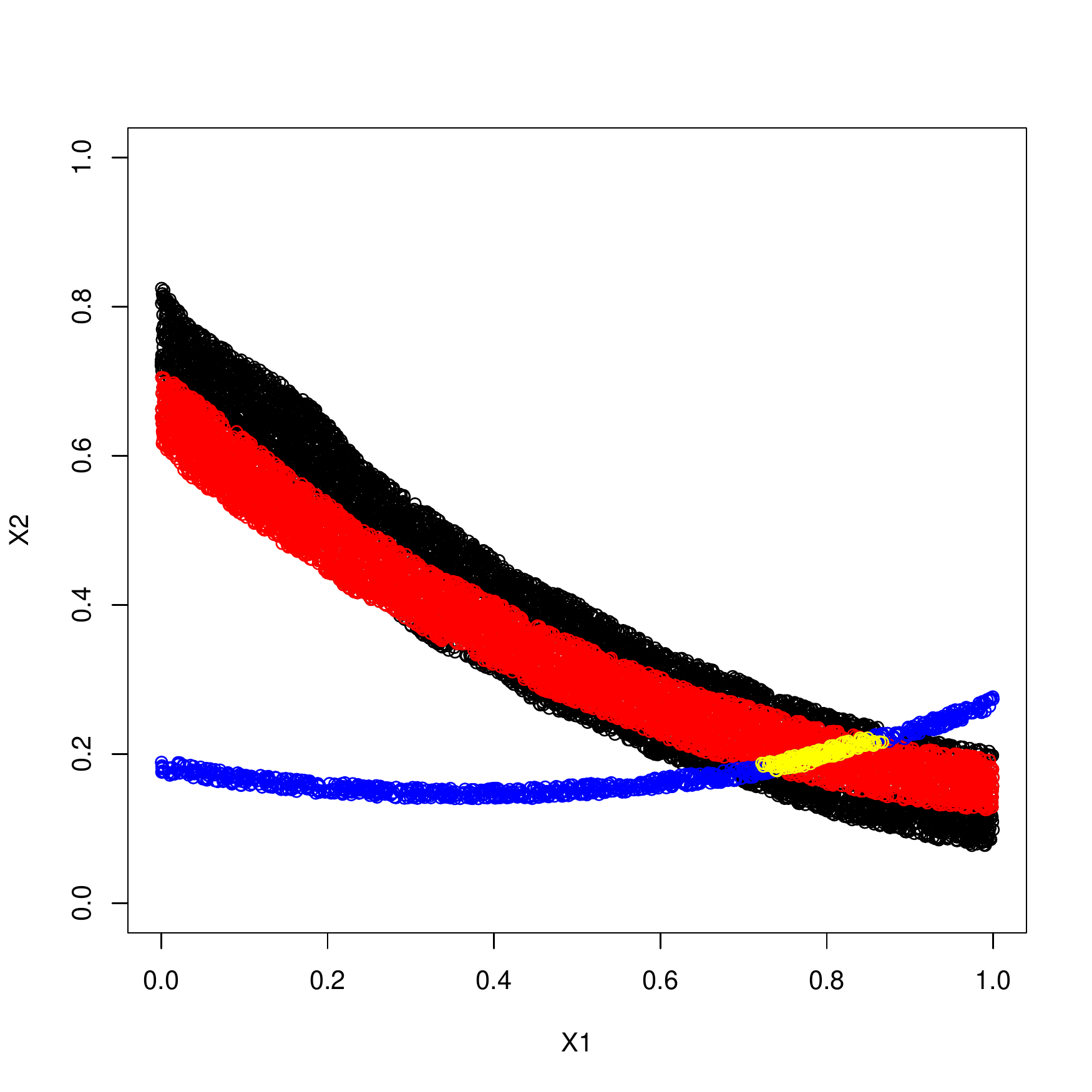}}
	\subfigure[DPS-145-37-132]{\includegraphics[scale=0.35]{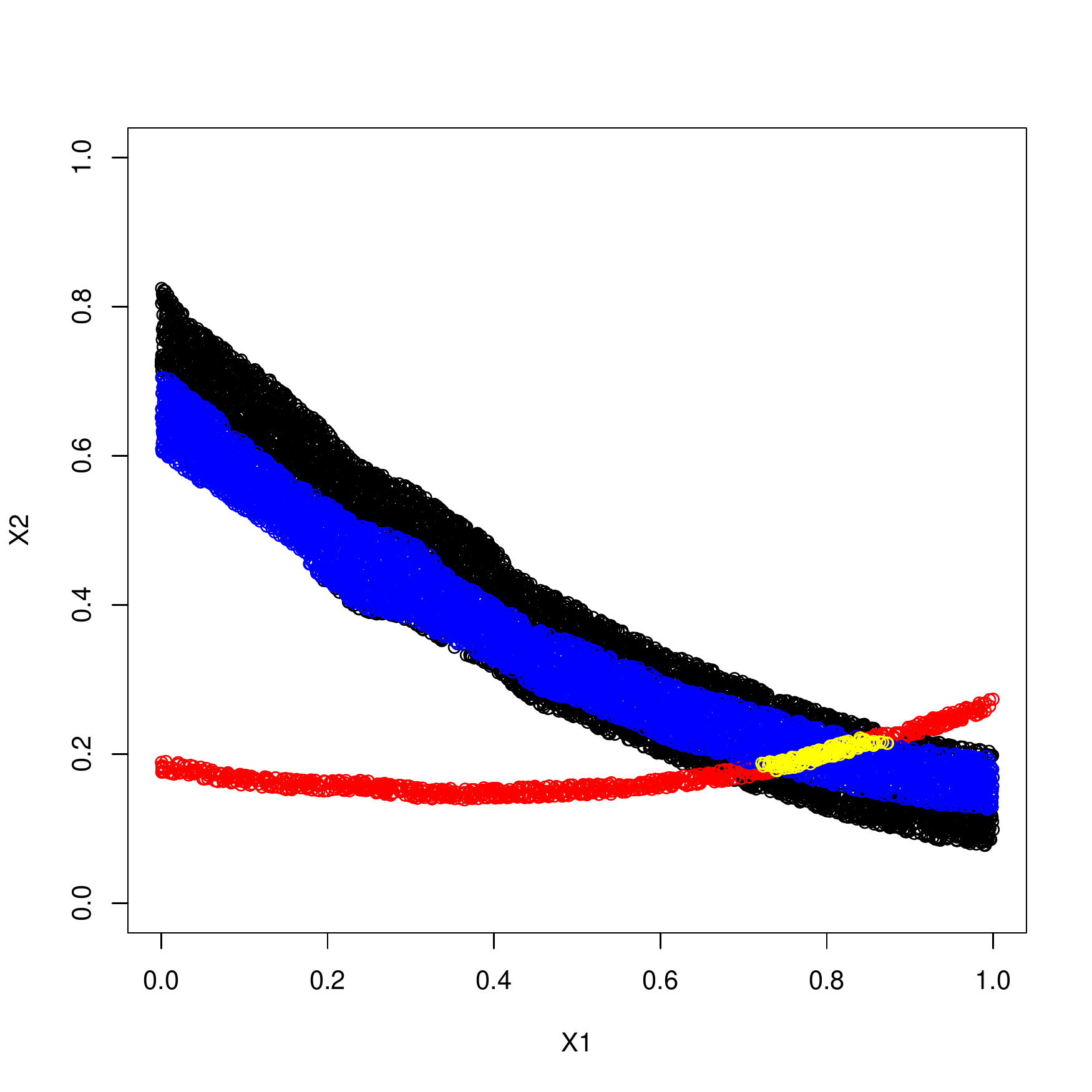}}
	\caption{Easom Function: Comparison of the inverse solution estimate as per the proposed MSCE method, with $n_0=15$ and $N=50$ points.} 
	\label{Fig: easom-dps-order}
\end{figure}

{\emph{Remark 2: Sensitivity of the order of DPS:}}  In principle (i.e., theoretically), if all scalar-valued inverse problems have been solved accurately, then the overall inverse solution of the simulator with {time series} response should also be estimated with high accuracy. However, in practice, it may be tempting to think that the order in which the three (in general, $k$) scalar-valued inverse problems are solved may affect the accuracy of the overall inverse problem for the underlying {time series} valued simulator.  Our investigations based on the simulated examples considered in this paper show that the order does not play a significant role. For instance, Figure~\ref{Fig: easom-dps-order} depicts the sensitivity of the order of DPS in the sequential contour estimation approach for the Easom function example. In Figure~\ref{Fig: easom-dps-order}, the point clouds represent $S_1, S_2, S_3$ and $\cap_{i=1}^3S_j$ in the order of black, red, blue and yellow for different DPS sequences shown in the figure captions.

Of course, this demonstration based on finitely many examples does not guarantee that the order will not matter for every MSCE implementation of an inverse problem for {time series} valued simulators. If the fitted surrogates at each $t_i^*$ are adequate to find the true inverse solution, then clearly the ultimate inverse solution found at the end should not differ.

\textbf{Extraction of the overall inverse solution $S_0$:} We approximate $S_0$ with $\cap_{j=1}^k S_j$ -- the intersection of $k$ scalar-valued inverse solutions obtained at the discretization-point-set (DPS). If there exists a solution of the underlying inverse problem for the dynamic ({time series} valued) simulator, then $\cap_{j=1}^k S_j$ will be nonempty. The following result establishes the existence of the inverse solution as per the proposed approach.

\begin{thm}
	Let $S_0 = \{x \in \chi: g(x, t_j) = g_0(t_j), j=1,2,...,L\}$ be the true inverse solution for a {time series} valued simulator $g(x)$ with respect to $g_0$, and $S_j = \{x \in \chi: g(x, t_j^*) = g_0(t_j^*)\}$  be the inverse solution at the $j$-th DPS point $t_j^*$, then, $S_0 \subseteq \cap_{j=1}^k S_j$.
\end{thm}

\textbf{Corollary~1:}\emph{ If $\cap_{j=1}^k S_j$ represents a single cluster, then $S_0$ is unique.} \\

\textbf{Practical Implementation:} \emph{If $\cap_{j=1}^k S_j$ generates multiple distinct clusters, then either the underlying inverse solution has multiple inverse solutions or we have detected some false solution along with the correct solution. This can be further ascertained by increasing the size of DPS and follow the proposed approach.}

We have omitted the proof of Theorem~1, as it is straightforward and not giving additional insights to this discussion.

The desired inverse solution would be 
\begin{eqnarray*}
\hat{x}_{opt} = argmin\{\|g(x)-g_0\|, x \in \cap_{j=1}^kS_j\}.
\end{eqnarray*}
To implement this, we obtain $k$ GP surrogates $\hat{g}(x, t_j^*)$,  after the final iteration, using all $N$ training points found in the due process of estimating $k$ contours.  Instead of the exact match, we accept the approximate inverse solutions as  $S_j(\delta) = \{x: |\hat{g}(x, t_{j}^*) - g_{0}(t_j^*)| < \delta\}$ for some small $\delta$, for each time point $t_j^*$ in DPS. This accounts for the round off errors and other approximations made during the implementation. This tolerance $\delta$ has to be judiciously chosen to accurately estimate the inverse solution set.

\begin{figure}[h!]\centering
	\includegraphics[scale=0.65]{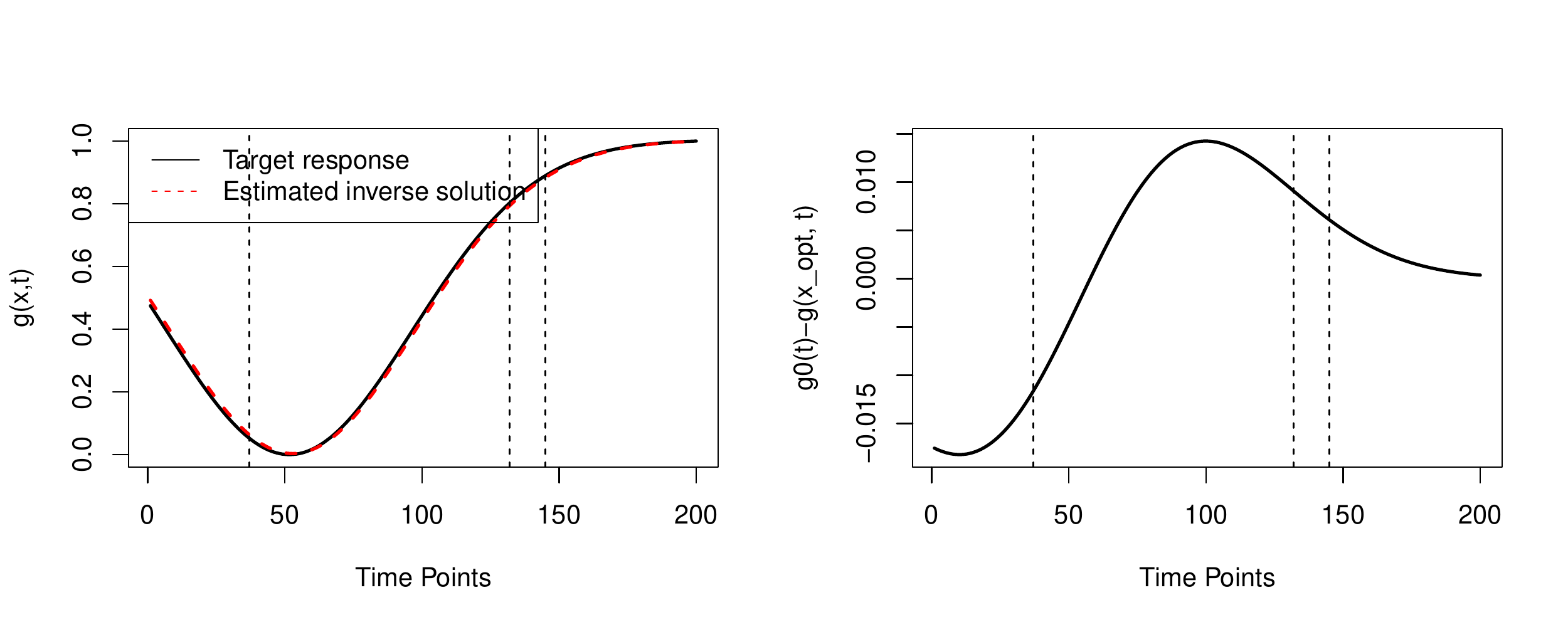}  
	\caption{Easom function: The left panel shows the target response in black solid line and the simulator response at $x_{opt}$ is shown by the dashed red curve. The right panel presents the difference $g_0(t)-g(\hat{x}_{opt},t)$.}
	\label{fig:easom-final-inverse-sol}
\end{figure}

For Easom function example, it is clear from Figures~\ref{fig:easom-seq-3-DPS-points} and \ref{Fig: easom-dps-order}, that the three contours intersect on a common point. Here, we set $\delta = 10^{-5}$, and the final inverse solution obtained is $\hat{x}_{opt} = (0.8188, 0.2029)$. Figure~\ref{fig:easom-final-inverse-sol} shows that the simulator response at $\hat{x}_{opt}$  (blue dashed curve) is virtually indistinguishable as compared to the target response (black solid curve).

{\small
	\begin{algorithm}[h!]
		\SetKwInOut{Input}{Input}\SetKwInOut{Output}{Output}
		\DontPrintSemicolon
		\Input{(1) Input parameters: $d,L,n_0, N$ \\ (2) {time series} valued computer simulator: $\{g(x,t_j), j=1,...,L\}$ \\ (3) Target response: $\{g_0(t_j),j=1,...,L\}$\\ (4) Tolerance: $\delta$ }
		\Output{(1) Final training set: $\texttt{xx}_{N\times d}$ and $\texttt{yy}_{L\times N}$\\ (2) Estimated inverse solution: $\hat{\texttt{x}}_{opt}$}
		\hrule
		Construct a DPS of size $k (\ll L)$ that would capture the important features of the target {time series} response, say, $(t_1^*, t_2^*, ..., t_k^*)$.	See Section~3 for the proposed regression spline based methodology. \;
		Choose $n_0$ points in $\chi=[0,1]^d$ using a maximum projection Latin hypercube design. Obtain the corresponding simulator response matrix $Y_{L\times n_0}$.\;
		\For{$j = 1, \ldots, {k}$}{
			Use contour estimation method for scalar-valued simulator to estimate $S_j(x)=\{x\in \chi \ : \ g(x,t_j^*) = g_0(t_j^*) \}$. Assume the size of initial design is $n_0 + (j-1)\cdot (N-n_0)/k$, whereas $(N-n_0)/k$ follow-up trials are added sequentially one at-a-time as per the EI criterion in Section~2.4. \;
			Augment the follow-up points to the initial design for the $(j+1)$-th scalar-valued inverse problem.\;
		}
		Fit final $k$ GP surrogates to $g(x, t_j^*)$ using all $N$ training points. Obtain $S_j = \{x: |\hat{g}(x, t_{j}^*) - g_{0}(t_j^*)| < \delta\}$ {and $U_j = \{x: |\hat{g}(x, t_{j}^*) - g_{0}(t_j^*)| < \alpha s(x,t_{j}^*)\}$} for $j=1,2,...,k$. \;
		Extract the final inverse solution as $\hat{x}_{opt} = argmin\{\|g(x)-g_0\|, x \in \cap_{j=1}^kS_j\}$, {and report the spread of $\cap_{j=1}^k U_j$ as the associated uncertainty measure}.\;
		\caption{Multiple scalar-valued contour estimation (MSCE) approach}
	\end{algorithm}
}

{This notion of extracting the inverse solution via $\cap_{j=1}^kS_j$ can be further extended to quantify the uncertainty in the inverse solution estimate. In spirit of the formulation of the improvement function in Section~2.4, define
	$$U_j = \{x \in \chi: |\hat{g}(x, t_j^*) - g_0(t_j^*)| < \alpha s(x, t_j^*)\}, \, j=1,2,...,k,$$
	where $\hat{g}$ is obtained from the final fit. Assuming $S_0$ is unique, and $\cap_{j=1}^k S_j$ is non-empty, the spread of $\cap_{j=1}^k U_j$ can be taken as a measure of uncertainty in estimating the inverse solution. It is intuitive to infer that the spread of $\cap_{j=1}^k U_j$ will converge to zero as the size of the training data increase to infinity. Here spread($\cap_{j=1}^k U_j$), later abbreviated as spread($U_j$) in Tables~\ref{Tab:results-examples} and~\ref{Tab: ms1}, is equal to $\sum_{r=1}^d Var(x_r)$, where $x_r$ is the vector of $r$th coordinate from the estimated inverse solution $\cap_{j=1}^k U_j$. Note that $U_j = S_j(\delta)$ for $\delta = \alpha s(x,t_j^*)$.} 

We summarize the key steps of the proposed {MSCE} approach in Algorithm~1.

\textbf{Approximate Inverse Solution:}
{
	If the simulator output and/or the target response are noisy then the exact match for the inverse solution would not exist, and hence the approximation using $\cap_{j=1}^k S_j(\delta)$ and $\cap_{j=1}^k U_j$ are viable options to obtain the closest possible inverse solution. The target response in our hydrological model is noisy (see Section~5). For a quick illustration, we introduce a random Gaussian noise term in the Easom simulator output (i.e., the time series response is $g(x,t)+\varepsilon(t)$) and the target response corresponds to the same $x_0 = (0.8, 0.2)$, $DPS =(145,37,132)$ and $n_0=15$ and $N=30$. Figure~\ref{fig:easom-noisy} presents the simulator responses corresponding to  $x \in \cap_{j=1}^k U_j$ and the best estimate of the inverse solution.

\begin{figure}[h!]\centering
	\includegraphics[scale=0.85]{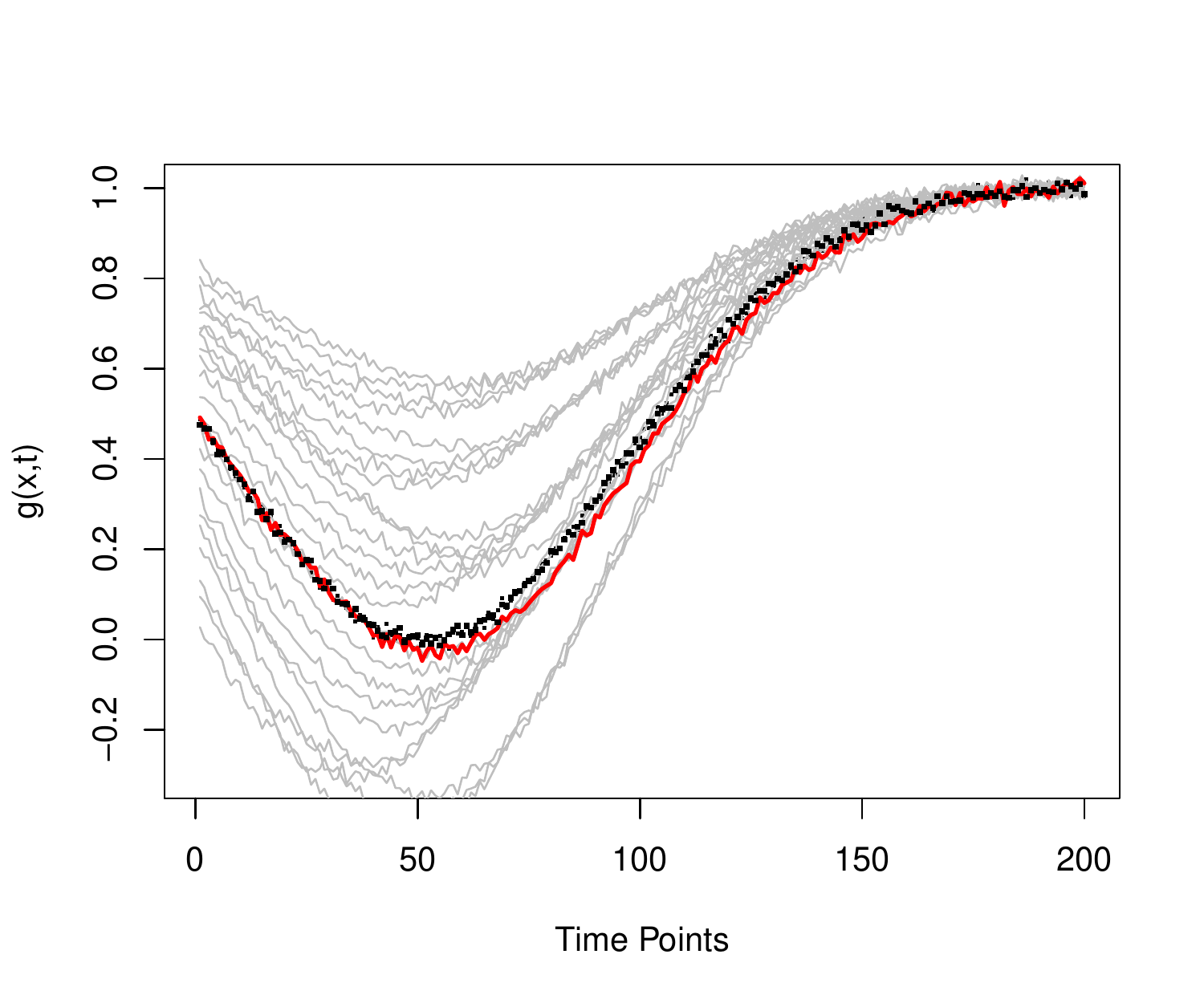}  
	\caption{Easom function: Black dots represent the target response, the red curve shows the best estimate of the inverse solution and the gray curves show the responses corresponding to $\cap_{j=1}^k U_j$.}
	\label{fig:easom-noisy}
\end{figure}

Given that the simulator returns noisy output, the final estimate appears to be a reasonably good approximation of the desired inverse solution. 
}

\section{Simulation studies}

In this section, we use three different test function based {time series} valued simulators to compare the performance of the proposed method with the modified history matching (HM) algorithm (Vernon et al., 2010; Bhattacharjee et al., 2019), the naive scalarization method (Ranjan et al., 2016), {and the saddlepoint approximation based EI (saEI) approach (Zhang et al. 2019)}. 
%
%
For performance comparison between the four methods, we use three popular goodness of fit measures called $R^2$, RMSE, normD, {and the uncertainty measure proposed  in Section~3 (i.e., the spread of $\cap_{j=1}^k U_j$)}. The objective would be to maximize $R^2$ and minimize RMSE, normD, and the spread of  $\cap_{j=1}^k U_j$. 

\begin{itemize}
	
	\item Root mean squared error given by
	\begin{eqnarray*}
		RMSE = \left(\frac{1}{L} \sum_{j=1}^{L} \mid g(\hat{x}_{opt}, t_j) - g_0(t_j) \mid ^2 \right)^{1/2}
	\end{eqnarray*}
	measures the discrepancy between the simulator response at the estimated inverse solution $\hat{x}_{opt}$ and the target response-series $g_0$.
	
	\item Coefficient of determination $R^2$ of the simple linear regression model fitted to the estimated inverse solution and the target response, i.e., $R^2$ of the following linear regression model:
	\begin{eqnarray*}
		g_0(t_j) = g(\hat{x}_{opt}, t_j) + \psi_j, j = 1, 2, \dots, L, 
	\end{eqnarray*}
	with the assumption of i.i.d. error $\psi_j$.

	\item Normalized discrepancy (on log-scale), between the simulator response at the estimated inverse solution and the target response
	\begin{eqnarray*}
		normD=\log\left(\frac{\left\|{g_0}-{g}\left(\hat{{x}}_{opt}\right)\right\|_{2}^{2}} {\left\|{g_0}-\bar{g}_0 1_{L}\right\|_{2}^{2}}\right)
	\end{eqnarray*}
	where $\bar{g}_0 = \sum_{t=1}^{L}g_0(t)/L$ and $1_{L}$ is an L-dimension vector of ones. Note that $1-\exp(normD)$ is a popular goodness of fit measure and often referred to as Nash–Sutcliffe Efficiency (Nash and Sutcliffe, 1970).
\end{itemize} 

%
%

{For each test function, 100 replications were run for different initial training data obtained via maxPro LHD (Joseph et al., 2015),  random test sets for optimizing the follow-up criteria, and candidate sets for extracting the inverse solutions.  We also prefixed the target series (consequently the DPS),  and the $n_0$ and $N$ combination. Note that the implementation of Scalarization method, saEI and MSCE requires sequential augmentation of one follow-up trial at-a-time by maximizing some criteria, and hence prefixing the initial design size ($n_0$) and a total budget ($N$) is in sync with these three methods. However, the (modified) HM approach accumulates follow-up points in batches and thus the total runsize will vary slightly with the initial design and/or the test sets.}

{
\subsection{Example 1: Easom Function (Michalewicz, 1996) contd.}
We begin by revisiting the illustrative example discussed in Section 3 to compare the inverse solutions arrived at by the four methods.  Recall that the initial design size is set to $n_0=15$, the total runsize to $N=50$, the target series corresponds to $x_0=(0.8,0.2)$ which led to $DPS = (145, 37, 132)$. Figure~\ref{Fig: easom1}  compares the three goodness of fit (GOF) measures ($R^2$, RMSE and norm-D) for all four methods over different replications.

\begin{figure}[h!]\centering
\includegraphics[scale= 0.65]{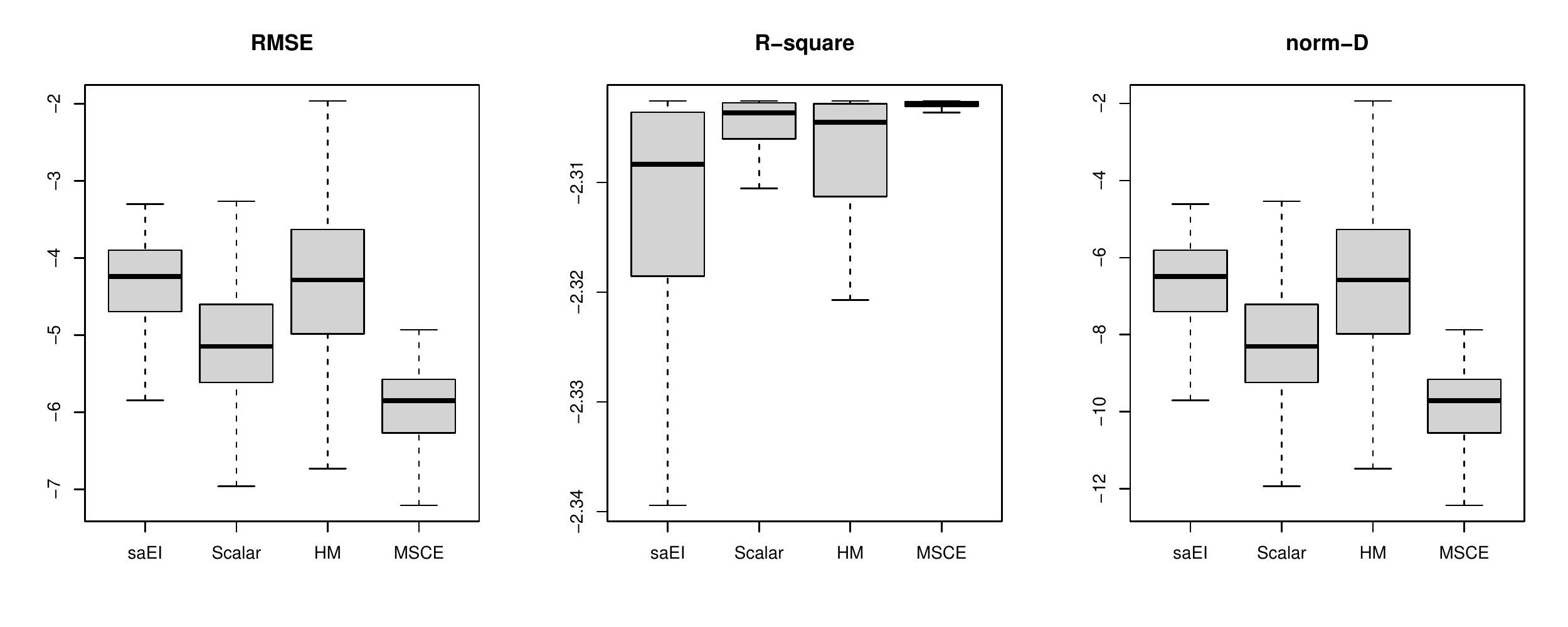}
\caption{Easom Function: GOF comparison between the saEI method, scalarization method, modified HM algorithm, and the proposed MSCE approach.} 
	\label{Fig: easom1}
\end{figure}

For better visual comparison, we have depicted the distributions of $\log(RMSE)$, $\log(normD)$ and $\log(R^2-0.9)$. It is evident from Figure~\ref{Fig: easom1} that MSCE outperforms the other competitors by a big margin with respect to all GOF measures.  We also compared the accuracy of estimated inverse solutions over the replications (see Figure~\ref{Fig: easom2}).

\begin{figure}[h!]\centering
	\includegraphics[scale= 0.6]{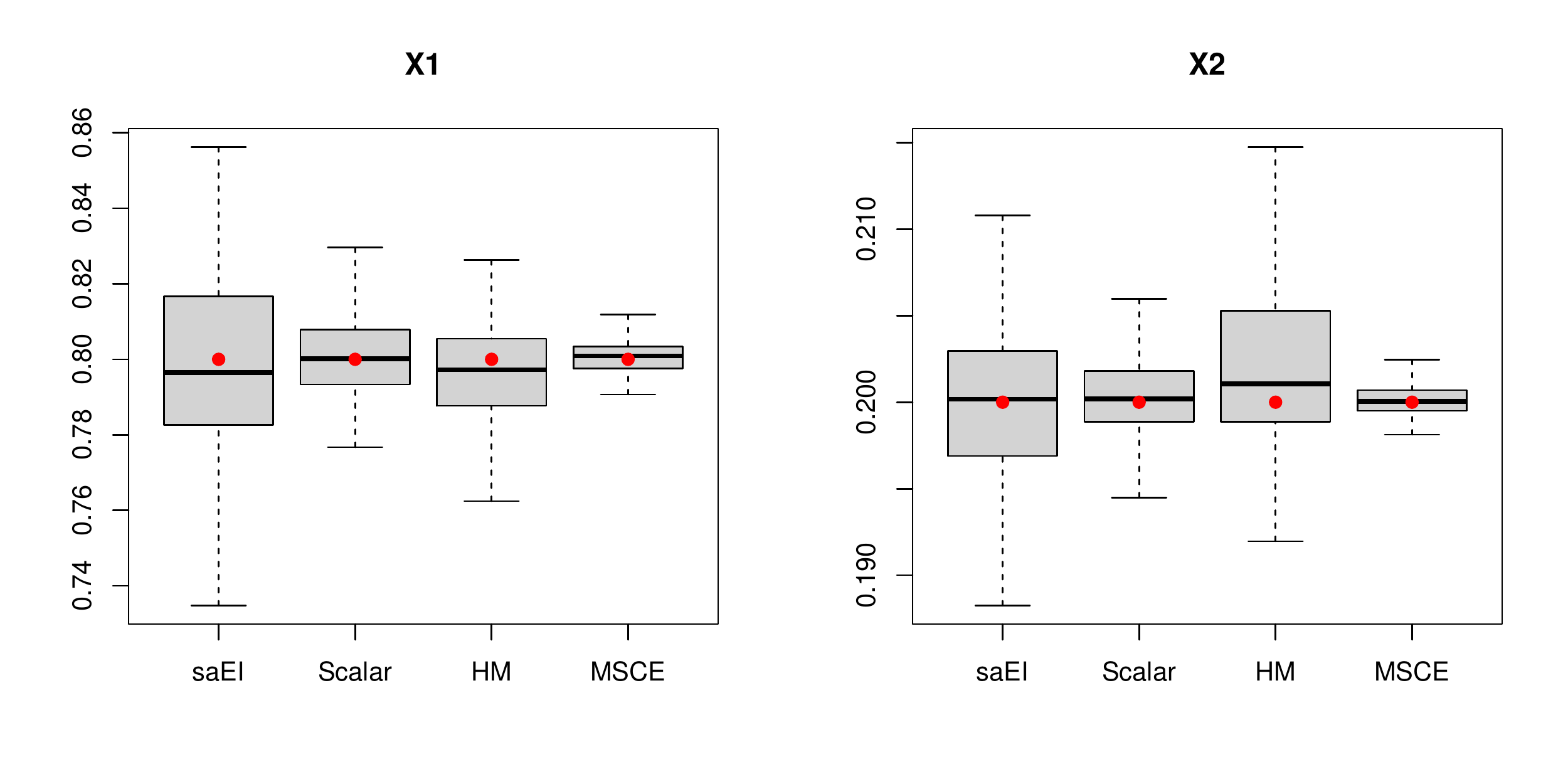}
	\caption{Easom Function: Distribution of $\hat{x}_{opt}$: comparison between the saEI, scalarization method, HM algorithm, and the MSCE approach. Red dot represents the true $x_0$ that generated $g_0$.} 
	\label{Fig: easom2}
\end{figure}

The left and right panels of Figure~\ref{Fig: easom2} shows the boxplots of $\hat{x}_{opt}$ for $x_1$ and $x_2$ respectively, for the four competing methods. The larger the boxplots, the bigger the uncertainties associated with the corresponding methods. Figure~\ref{Fig: easom2} shows that all methods are able to estimate the inverse solution, but the proposed method MSCE does it more accurately (i.e., the variation is smallest around the true value) as compared to the other competitors. As an alternative means of  uncertainty quantification (UQ), we computed the total dispersion of $\cap_{j=1}^k U_j$ for all methods in different replications and the results are summarized in Figure~\ref{Fig: easom3}. The lower the boxplots are located on the $y$-axis, the better the methods perform.


\begin{figure}[h!]\centering
	\includegraphics[scale= 0.6]{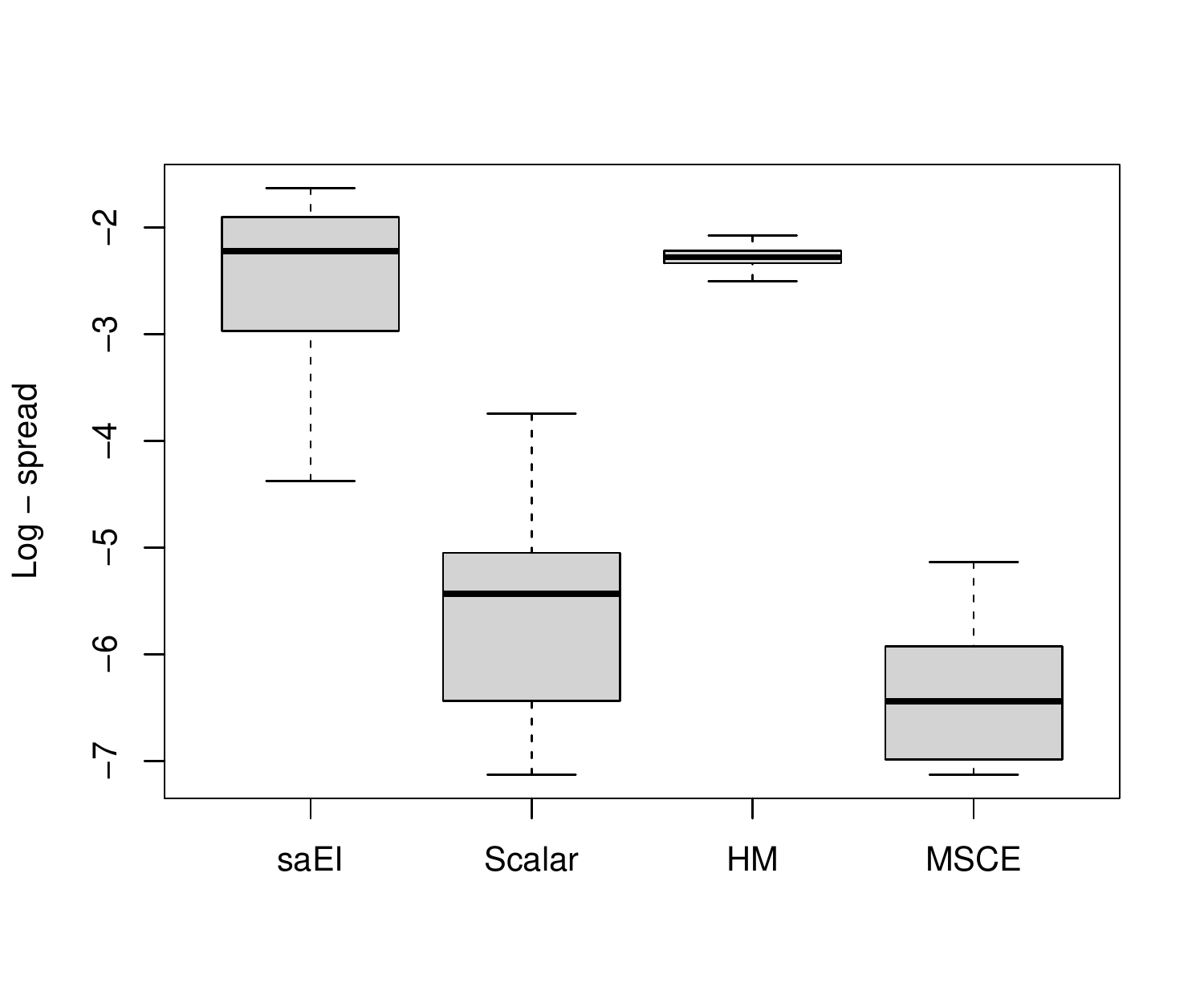}
	\caption{Easom Function: Distribution of the log-spread of $\cap_{j=1}^k U_j$- UQ in the estimate of the inverse solution. Comparison between the saEI, scalarization method, HM algorithm, and the MSCE approach. } 
	\label{Fig: easom3}
\end{figure}

As per this UQ measure as well, we can see that the proposed method gives the most accurate results. Interestingly, the HM method gives consistently inaccurate results. 

\bigskip
\subsection{More test function based examples}
In this section, we compare the performance of the four methods via several test function based {time series} valued computer simulators. All results are averaged over 100 replications. The test functions are listed as follows:

\begin{enumerate}
		\item Levy function: The original Levy function by Laguna and Marti (2002) produces scalar response for an arbitrary input dimension $d$. We have modified the test function to generate {time series} outputs. For $d=2$, let 	%
	\begin{eqnarray*}
	 y_t(x) &=&  sin(\pi t)^2 + \left[\left(\frac{t}{5}-1\right)^2(1+10(sin(0.5\pi t+1))^2)\right] * \\
	 && [(w_1-1)^2(1+10(sin(\pi w_1+1))^2)]+
	(w_2-1)^2(1+(sin(2 \pi w_2))^2) ,
\end{eqnarray*}
	where $w_i = 1+(x_i-1)/4$, and $x_i \in (-10,10)$. For the simulation study in this paper, we have fixed $n_0=15, N=45$, $x_0 = (0.5, 0.5)$ and $DPS = (40, 110, 170)$.

	\item Harari and Steinberg (2014): The simulator takes $d(=3)$-dimensional inputs $x=(x_1, x_2, x_3) \in [0,1]^3$ and produces {time series} response as per
	$$y_t(x)= \exp(3x_1t + t)\times  \cos(6x_2t+ 2t - 8x_3 - 6)$$
	where $t \in [0,1]$ on a 200-point equidistant grid. We assumed $x_0 = (0.522, 0.95,0.427)$ (drawn randomly) for generating the target series and found $DPS = (118, 26, 95)$ as per the algorithm outlined in Section~3. Furthermore, the simulation study was conducted with the initial design size of $n_0=20$ and a total budget of $N=50$.
	
	\item Bliznyuk et al. (2008) presents an environmental model which simulates a pollutant spill caused by a chemical accident. Here, the input space is $x = (x_1, x_2, x_3, x_4, x_5)^T \in [7,13] \times [0.02, 0.12] \times [0.01, 3] \times [30.01, 30.304] \times [0,3]$, and the simulator outputs are generated as:
	
		\begin{eqnarray*}
		y_t(x) &=& \frac{x_1}{\sqrt{x_2t}} \exp\bigg(\frac{-x_5^2}{4x_2t}\bigg)  +  \frac{x_1}{\sqrt{x_2(t - x_4)}}\exp\bigg(\frac{-(x_5-x_3)^2}{4x_2(t - x_4)}\bigg)I(x_4<t)  
	\end{eqnarray*}
	with $t \in [35.3, 95]$ defined over a 200 point equidistant grid. The target {time series} response corresponds to $x_0 = (9.640, 0.059, 1.445, 30.277, 2.520)^T$ (randomly chosen).
	The corresponding DPS turns out to be $(30, 7, 61, 14)$ and the simulation study assumed
	$n_0 = 30$ and $N=90$.
	
\end{enumerate}

	\begin{table}[h!]
	\centering
	\caption{Performance comparison of the four methods (saEI, Scalarization, HM and the proposed MSCE) with respect to four goodness of fit measures: $R^2$, RMSE, norm-D and spread of $\cap_{j=1}^kU_j$ (presented in log-scale to highlight the difference). The numbers in parentheses denote the standard error, and the red-bold values show the best among the four methods.} 
	\medskip
		\text{(Modified) Levy function $(d=2)$}\\
	\begin{tabular}{llcccccc}
		\hline
		\bf{Methods} &&& \bf{saEI} & \bf{Scalar} & \bf{HM} & \bf{MSCE}\\
		\hline
\bf{spread($U_j$)}&&& -2.1 (0.15) & -2.5 (0.37) & -2.1 (0.27) & \textcolor{red}{\bf -3.3 (0.19)}\\
\bf{RMSE} 	   &&& -8.4 (1.3) & -10.2 (0.88) & -10.7 (0.92) & \textcolor{red}{\bf -12.0 (0.59)}\\
\bf{R-squared} &&& -0.07 (0.13) & -0.003 (0.014) & -0.003 (0.016) & \textcolor{red}{\bf -3$\times 10^{-5}$ (6$\times 10^{-5}$)}\\
\bf{norm-D}    &&& -1.1 (2.7) & -4.7 (1.76) & -5.76 (1.84) & \textcolor{red}{\bf -8.9 (1.2)}\\
		\hline
	\end{tabular}
	\\[12pt]  \text{Harari and Steinberg (2014) function $(d=3)$}\\
	\begin{tabular}{llcccccc}
		\hline
		\bf{Methods} &&& \bf{saEI} & \bf{Scalar} & \bf{HM} & \bf{MSCE}\\
		\hline
\bf{spread($U_j$)}  &&& -2.2 (0.73) & -2.02 (0.76) & -1.73 (0.40) & \textcolor{red}{\bf -3.2 (0.91)}\\
\bf{RMSE} 		 &&& -4.4 (0.53) & -5.23 (0.60) & -4.76 (0.72) & \textcolor{red}{\bf -6.1 (0.41)}\\
\bf{R-squared} 	 &&& -1.2 (0.37) & -0.77 (0.10) & -0.86 (0.32) & \textcolor{red}{\bf -0.7 (0.005)}\\
\bf{norm-D} 	 &&& -2.0 (1.05) & -3.68 (1.19) & -2.74 (1.45) & \textcolor{red}{\bf -5.5 (0.82)}\\
		\hline
	\end{tabular}
	\\[12pt] \text{Bliznyuk et al. (2008) function $(d=5)$}\\
\begin{tabular}{llcccccc}
	\hline
	\bf{Methods} &&& \bf{saEI} & \bf{Scalar} & \bf{HM} & \bf{MSCE}\\
	\hline
\bf{spread($U_j$)} &&& -0.7 (0.027) & -0.79 (0.05) & -0.61 (0.027) & \textcolor{red}{\bf -0.85 (0.026)}\\
\bf{RMSE} 		&&& -5.8 (0.52) & -5.77 (0.47) & -6.55 (0.51) & \textcolor{red}{\bf -7.47 (0.419)}\\
\bf{R-squared} 	&&& -0.7 (0.019) & -0.72 (0.044) & -0.70 (0.0052) & \textcolor{red}{\bf -0.69 (0.0006)}\\
\bf{norm-D} 	&&& -3.8 (1.04) & -3.73 (0.94) & -5.29 (1.02) & \textcolor{red}{\bf -7.17 (0.84)}\\
	\hline
\end{tabular}
	\label{Tab:results-examples}
\end{table}

}

{
It is clear from Figures~\ref{Fig: easom1} -- \ref{Fig: easom3} and Table~\ref{Tab:results-examples} that the proposed MSCE method significantly outperforms the three competitors (saEI, Scalar and HM methods) with respect to all four goodness of fit (GOF) criteria for four test function based simulators ranging from $d=2$ to $d=5$. Once again recall that the objective is to maximize $R^2$ whereas minimize the other three GOF measures. 

}


\section{Real Application: Rainfall-Runoff Example} 
\begin{figure}[h!]
	\begin{center}
		\begin{tabular}{cc}
			\includegraphics[scale=0.9]{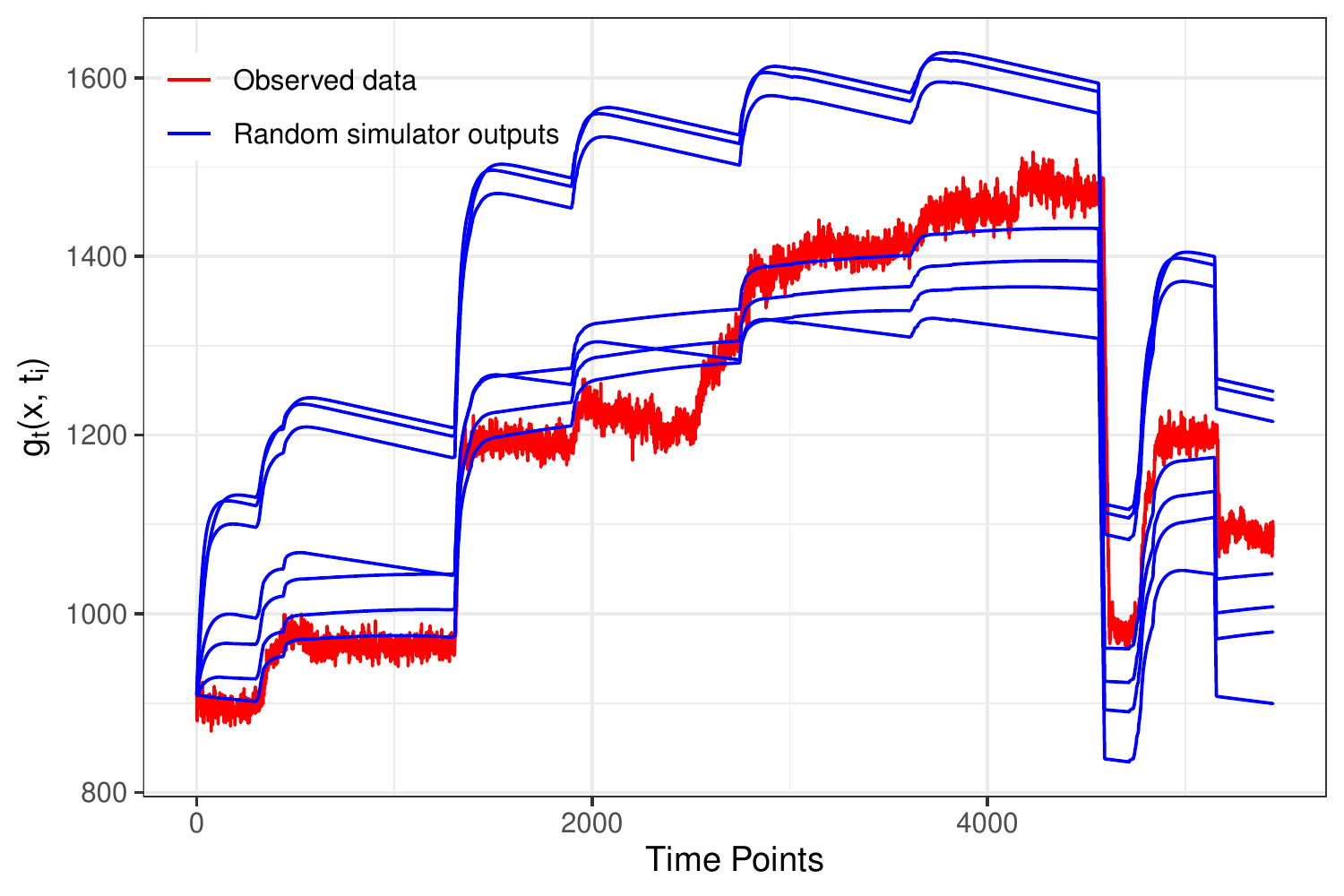} 
		\end{tabular}
		\caption{Hydrological model: Target response $g_0$ along with a few random simulator outputs observed over 5445 time-points.}
		\label{ms_random}
	\end{center}
\end{figure}

The motivating application in Bhattacharjee et al. (2019) used a hydrological simulator $-$ Matlab-Simulink model introduced by Duncan et al. (2013)  $-$ to study the rainfall-runoff relationship for a windrow composting pad. The following four parameters have been identified as the inputs with most significant influence on the output: depth of surface, depth of sub-surface and two coefficients of the saturated hydraulic conductivity ($K_{sat1}$ and $K_{sat2}$). Interested readers can see Duncan et al. (2013) for further details on the hydrological model. For the inverse problem, the target response is the rainfall-runoff data ($g_0$) observed from the Bioconversion center at the University of Georgia, Athens, USA. Figure~\ref{ms_random} depicts the observed target response and a few random outputs from the hydrological model.

\begin{figure}[h!]
	\begin{center}
		\begin{tabular}{cc}
			\includegraphics[scale=0.75]{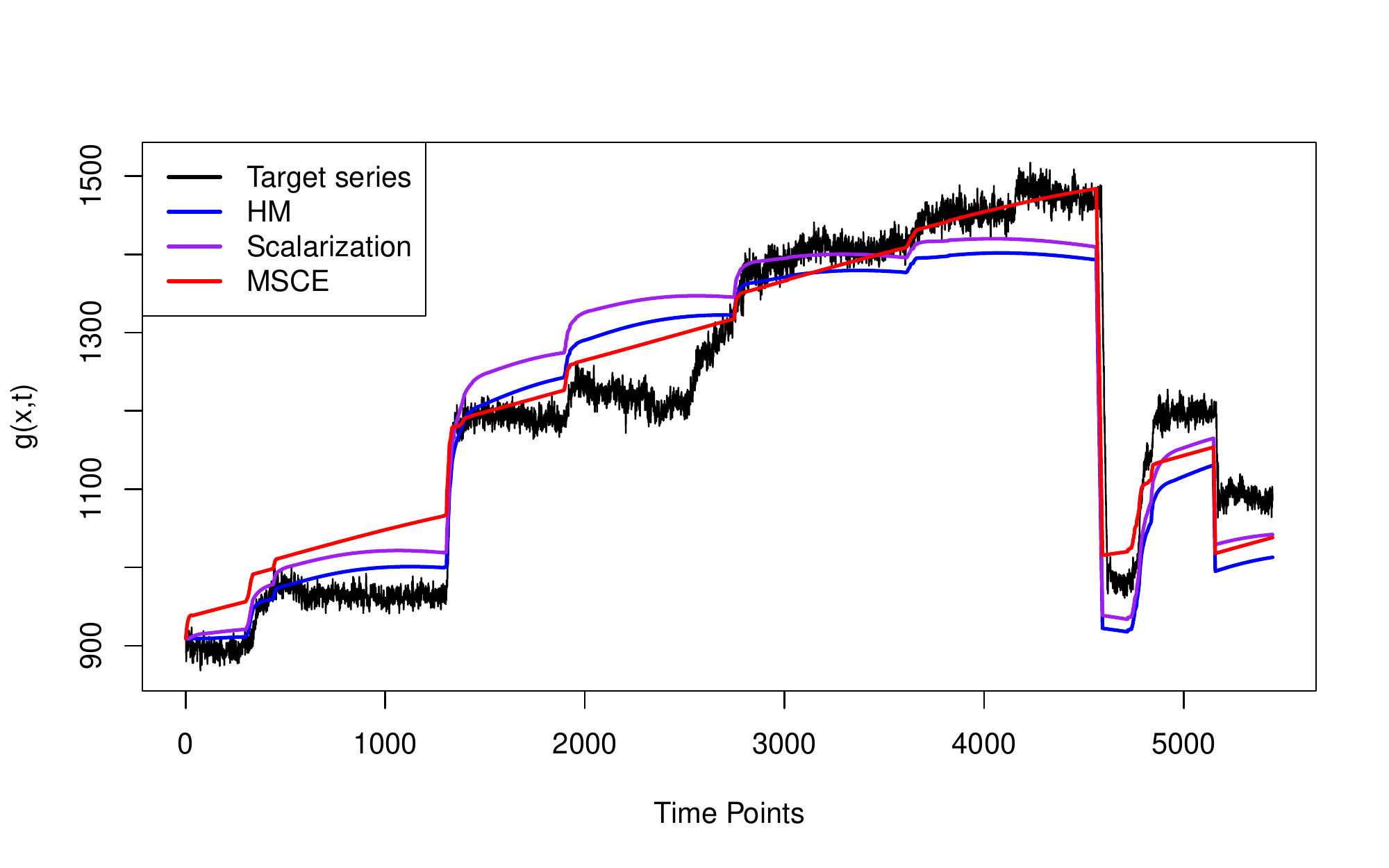} 
		\end{tabular}
		\caption{Hydrological model: {Black curve shows the target response, and the estimated inverse solutions corresponding to the modified HM method is shown in blue, MSCE by the red curve, and scalarization method by the purple curve.}}
		\label{Figs:ms_a}
	\end{center}
\end{figure}

{It is clear from Figure~\ref{ms_random} that the target response appears to be noisier than the simulator response, and a little biased as well.} {The optimal knots in the regression spline approximation of the target response led to $DPS=\{4557, 3359, 4702, 4085\}$. {It is important to observe that three of the time-points in DPS are between $t=4000$ and $t=5000$ -- the region with a big sudden dip. Although this clustering behaviour is different from the earlier examples, it may be expected as this drastic change in the nature of the target series overpowers small variations in the other region.} We implement the proposed MSCE approach with $n_0 = 40$-point maxPro Latin hypercube design as an initial design and added additional 10 follow-up points. The results are compared with the modified HM approach and the scalarization method. The saEI approach could not be implemented here, because the R package DynamicGP required passing the computer simulator function, which we did not have access to in the required format. Table~\ref{Tab: ms1} {summarizes} the GOF results.

\begin{table}[h!]
	\centering
	\caption{Hydrological model: Goodness-of-fit comparisons of the proposed MSCE methods with the modified HM and scalarization methods.} 
	\medskip
	\begin{tabular}{lcccc}
		\hline
		\bf{Methods} & \bf{Scalar} & \bf{HM} & \bf{MSCE}\\
		\hline
		\bf{spread($U_j$)}& 0.3053 & 0.2892 & \textcolor{red}{\bf 0.2860}\\
		\bf{RMSE}      & 67.06  & 64.69 & \textcolor{red}{\bf 53.96}\\
		\bf{R-squared} & 0.8824 & 0.8888 & \textcolor{red}{\bf 0.9314}\\
		\bf{norm-D}    & 0.1225 & 0.1140 & \textcolor{red}{\bf 0.0793}\\
		\hline
	\end{tabular}
	\label{Tab: ms1}
\end{table}

The results shown in Table~\ref{Tab: ms1} are consistent with the trends from the test function based simulators (in Table~\ref{Tab:results-examples}). That is, the proposed MSCE approach outperforms the other competitors in terms of finding the closest match for the observed runoff data with respect to all four metrics. This is also evident from the visual comparison (Figure~\ref{Figs:ms_a}) of the simulator responses corresponding to the estimated inverse solutions by different methods. 
}

\section{Concluding Remarks and Future Research}

In this paper, we have proposed a new {MSCE} approach of solving the inverse problem for {time series} valued computer simulators by first carefully selecting a handful of time-points for discretizing the target response series (called the DPS), and then iteratively solve multiple scalar-valued inverse problems at the DPS using the popular sequential algorithm via expected  improvement approach developed by Ranjan et al. (2008). The final inverse solution for the underlying dynamic simulator is obtained via the intersection of all scalarized inverse solutions. In this paper, we have suggested using a natural cubic spline based method for systematically finding the DPS. Based on the our simulation study using several test functions and a real-life hydrological simulator, it is clear that the proposed {MSCE} method outperforms three competing methods: scalarization technique (Ranjan et al., 2016), modified HM algorithm (Bhattacharjee et al., 2019) and saEI method (Zhang et al., 2019). {Although we do not have any theoretical justification yet, an intuitive explanation could be that (a) saEI uses saddlepoint approximation, which may not be very accurate;  (b) the scalarization method uses GP as a surrogate for the Euclidean distance between the target response and the simulator runs at all time-points, which becomes non-stationary around the inverse solutions, and hence could be a source of inaccuracy; (c) the two-fold modification of the original HM method adopted in this paper may have made it less efficient. In contrast, the proposed method carefully selects the DPS and then use one of the most efficient EI criterion for iteratively solving the  inverse problem.}

There are a few important remarks worth mentioning. 
%
(1) When finding an optimal DPS using spline-based technique, we followed a greedy ``forward variable selection" type approach and identified one best knot at-a-time. The ``best selection" type approach may lead to a better solution, however, it is computationally expensive (seemingly impractical) in finding the best DPS. 
%
(2) For solving the scalar-valued inverse problems at the $j$-th element of the DPS, we took the size of the initial design be $n_0 + (j-1)\cdot (N-n_0)/k$ and budget of follow-up points is $(N-n_0)/k$. Based on our preliminary simulation study, we found no significant improvement in accuracy by changing the order of DPS for solving the scalar-valued inverse problems. We divided the follow-up point resources $N-n_0$ equally among the $k$ scalar inverse problems, however, an efficient distribution of total budget $N$ can be further investigated. 
%
{
(3) A recent paper (Toscano-Palmerin and Frazier, 2022) proposes a new Bayesian methodology for solving the inverse problem for {time series} valued simulators. It would be interesting to compare the performance of our proposed frequentist MSCE approach with their Bayesian optimization technique with computationally expensive integrands.
(4) Since the responses are time series in nature, {one can investigate including time-correlation structure in the steps of MSCE, for instance, the surrogates at multiple $t^*_j$, for improved efficiency. }
%
(5) This paper assumes the existence of the inverse solution. Although the proposed methodology gives approximate solution in the presence of small noise, further research is required to find the best approximation of the inverse solution if it does not exist in the search space. 
}

%
%

\bigskip\noindent{\bf Conflict of interest statement:} On behalf of all authors, the corresponding author states that there is no conflict of interest.

\section*{Acknowledgements}
We would like to thank the Editor, the Guest Editor and the three referees for their helpful comments and suggestions which led to significant improvement of the paper.

\clearpage
\section*{Appendix A: Cost of Constructing Optimal DPS}

Suppose we need to construct the DPSs of size $j=1,2,...,k$, and the target series has been observed over 200 time points.  Then the costs of constructing these DPSs using the two methods are as follows.

\emph{Sequential search}: The first optimal knot can be found by fitting 200 different multiple linear regression (MLR) models with 4+1 regression coefficients each (4 for the cubic polynomial and 1 for the knot location term) and then comparing the goodness of fit criterion (e.g., MSE or $R^2_{adj}$). The second optimal knot, given the first one is already known, can be found by fitting 199 different MLR models with 4+2 coefficients each, and so on. That is, in total, for sequentially finding $k$ optimal knots using this method, one needs to fit 
$\sum_{j=0}^{k-1} (200-j) = 200k - k(k-1)/2$
different MLR models. In terms of computational complexity, the total cost would be 
$$\sum_{j=1}^{k} (200-(j-1))\cdot O((4+j)^3),$$
where $O((4+j)^3$ represents the computational cost of fitting a cubic-spline regression model to the target series with $j$ knots. 

\emph{Simultaneous search}: Here, the cost is heavily controlled by the resolution of the search grid, and how exhaustive the search is. For consistency, we find the one-knot optimal set in the exact same manner as in the ``sequential search" method, i.e., search the optimal knot over a 200-point grid. If we use the same 200-point grid, then we would have to fit ${200 \choose j}$ MLR models for finding optimal DPS with $j$ knots. That is, the total cost of constructing optimal DPS sets of size $j=1,2,...,k$ would be
$$\sum_{j=1}^{k} {200 \choose j}  \cdot O((4+j)^3).$$
Since ${200 \choose j}$ grows very rapidly with $j$, we follow a computationally cheaper approximation and randomly selected $200\cdot j$ candidate points for estimating the optimal DPS of size $j$. This is clearly much greater than the cost associated with the sequential search method.

Undoubtedly, the sequential search method does not guarantee the global optimum, and may give a sub-optimal estimate of the DPS. However,the sequential method will eventually iterate through all time-points, the and accuracy of DPS will increase to the maximum achievable level. 














\end{document}